\documentclass[a4paper,fleqn,usenatbib]{mnras}

\usepackage{newtxtext,newtxmath}

\usepackage[T1]{fontenc}
\usepackage{ae,aecompl}

\usepackage{graphicx}	
\usepackage{amsmath}	
\usepackage{amssymb}	
\usepackage{mathtools}
\usepackage{multirow}

\title[Black Holes Stop Host Galaxy Growth]{How Black Holes Stop Their Host Galaxy from Growing Without AGN Feedback}

\author[Eastwood \& Khochfar]{
Daniel S. Eastwood$^{1}$\thanks{E-mail: deastw@roe.ac.uk (DSE)}
and Sadegh Khochfar$^{1}$
\\
$^{1}$Institute for Astronomy, University of Edinburgh, Royal Observatory, Edinburgh EH9 3HJ, UK
}

\date{Accepted XXX. Received YYY; in original form ZZZ}

\pubyear{2018}

\DeclarePairedDelimiter\abs{\lvert}{\rvert}%
\begin{document}
\label{firstpage}
\pagerange{\pageref{firstpage}--\pageref{lastpage}}
\maketitle

\begin{abstract}
Super-massive black holes (SMBHs) with $M_{\bullet} \sim 10^9 $ M$_{\odot}$ at $z>6$ likely originate from massive seed black holes (BHs). We investigate the consequences of seeding SMBHs with direct collapse BHs (DCBHs) ($M_{\bullet}=10^{4-6}\, \mathrm{M}_\odot$) on proto-galactic disc growth. We show that even in the absence of direct feedback effects, the growth of seed BHs reduces the development of gravitational instabilities in host galaxy discs, suppressing star formation and confining stars to a narrow ring in the disc and leading to galaxies at $z \sim 6$ which lie above the local BH-stellar mass relation. The relative magnitude of cosmic and BH accretion rates governs the evolution of the BH-stellar mass relation. For typical DCBH formation epochs, $z_{\rm{i}} \sim 10$, we find star formation is inhibited in haloes growing at the average rate predicted by $\Lambda$CDM which host BHs capable of reaching $M_{\bullet}\sim 10^9 \, \mathrm{M}_{\odot}$ by $z\gtrsim6$. Slower growing BHs cause a delay in the onset of star formation; a $M_{\bullet} \sim 10^6 $ M$_{\odot}$ seed growing at $0.25$ times the Eddington limit will delay star formation by $\sim100$ Myr. This delay is reduced by a factor of $\sim10$ if the halo growth rate is increased by $\sim 0.6\, \sigma$. Our results suggest that SMBHs seeded by DCBHs and their host galaxies form in separate progenitor haloes. In the absence of subsequent mergers, higher than average cosmic accretion or earlier seed formation ($z_{\rm i} \sim 20$) are required to place the evolving BH on the local BH-stellar mass relation by $z=6$.
\end{abstract}




\section{Introduction} \label{introduction}

The relationship between super-massive black holes (SMBHs) and their host galaxies is an active area of research \citep[see, e.g.][]{Kormendy2013, Hickox2014, Delvecchio2015, Bongiorno2016, Yang2018}. Several empirical correlations between the  mass of a black hole (BH) and the physical properties of its host galaxy have been reported \citep[e.g.][]{Magorrian1998, Merloni2003, Merritt2006, Kormendy2009}. Of these correlations the BH mass-stellar velocity dispersion  (the $\mathrm{M_{\bullet}} - \sigma$ relation) \citep[e.g][]{Ferrarese2000, Gebhardt2000, Mcconnell2011} historically gave the first clues on a feedback driven co-evolution of BHs and their host galaxies \citep{Silk1998, King2003}.

Theoretical and observational studies suggest that major mergers play a fundamental role in establishing feedback and feeding cycles \citep[e.g.][]{DiMatteo2005, Hopkins2006, Kormendy2011}. Further support for the importance of mergers comes from the increased scatter in the BH-host correlations at larger redshifts which is a natural consequence of the central-limit theorem and an increasing number of BH mergers for SMBH toward low redshifts \citep{Schawinski2006, Peng2006, Hirschmann2010}. However, mergers are not the only physical processes involved; offsets in the $\mathrm{M_{\bullet}}-\mathrm{M_{Bulge}}$ relation corresponding to disc galaxies can be explained through the co-evolution of SMBHs with their disc-galaxy hosts through secular processes \citep{Volonteri2016, Simmons2017, Martin2018}.

SMBHs with masses of $M_{\bullet}\sim 10^{9}\, \mathrm{M_{\odot}}$ have been observed in galaxies at  high redshifts ($z\sim$ 6 -- 7) \citep{Fan2006a, Mortlock2011, Banados2018} corresponding to less than a gigayear after The Big Bang. This population of BHs must have had a rapid formation process to reach the masses observed at this early epoch. Indeed, if the growth rate of SMBHs is limited by the Eddington accretion rate \citep[see however,][]{Natarajan2012,Pacucci2017}, they must be seeded by some massive progenitor at an early epoch $z \geq 10$, prior to the onset of reionization and the shutdown of Population-III stars \citep{Paardekooper2015,Johnson2013a}.

Given the e-folding nature of an Eddington limited BH-growth rate on a Salpeter time scale of $t_{\mathrm{Sal}} = 0.45 \, \eta/(1 - \eta)\; \mathrm{Gyr}$ (where $\eta \sim 0.1$ is the radiative efficiency \citep[see, e.g.][]{King2008}), varying the initial seed mass by factors of ten can have a strong impact on relaxing the constraints on the formation time in the early Universe. Consequently, various seed formation processes are discussed \citep[see, e.g. for a review][]{Volonteri2010}, including population III stellar remnants \citep{Madau2001} and the collapse of dense stellar clusters \citep{Clark2008,Yajima2016} or the direct collapse of gas through the state of a super-massive star  \citep{Bromm2003,Lodato2006,Begelman2006,Begelman2010,Agarwal2012b}. The latter channel has received heightened attention due to the massive seeds it produces and the ability to grow to super-massive scales with less stringent constraints on the average accretion rate \citep{Agarwal2012b, Latif2013, Latif2014, Pacucci2015, Agarwal2016a}. \footnote{\citet{Johnson2011} simulated the radiative feedback from such a seed BH showing that the average accretion rate is very low, indicating that the feedback might off-set the advantage you gain of having a higher initial mass.}

Direct collapse BHs (DCBHs) form during the collapse of pristine gas in haloes with virial temperatures of $T_{\rm v}\gtrsim 10^{4}$ K \citep{Bromm2003}. Provided a halo remains pristine and the local intensity of the Lyman-Werner radiation field is greater than the critical intensity required to dissociate any H$_{2}$ gas \citep{Agarwal2016}, cooling within the halo will only take place via atomic hydrogen. The gas temperature in such a halo will be kept at $T_{\rm g}\sim 10^{4}$ K during collapse with a Jeans Mass of $M_{\rm J}\sim 10^{6} \, \mathrm{M_{\odot}}$; preventing the fragmentation into gas clumps and stars, and leading to the isothermal collapse of a massive gas cloud into a single BH \citep{Bromm2003}, possibly via an intermediate stage of a super-massive star \citep{Begelman2010}. This process results in the formation of massive seed BHs with masses of $M_{\bullet}\sim 10^{4}-10^{6}\, \mathrm{M_{\odot}}$ at $z\sim 10 - 20$, prior to the formation of the host galaxy in the halo \citep{Agarwal2012b}. If SMBHs are truly seeded by DCBHs it would affect the early stages of galaxy evolution. Gas build up around the gravitational potential well of a DCBH through cosmological accretion and halo merging, would not only lead to further BH growth but also potentially to the gradual growth of a proto-galaxy around the BH. Besides feedback from the BH affecting the proto-galaxy, initially such a proto-galaxy would be gravitationally dominated by the mass of the BH as well. However, it is not clear how this would affect the processes of galaxy evolution, such as star formation, and the cycle of baryons in the galaxy.

Recently, a first potential candidate for an observed DCBH system has been proposed (\citealt{Sobral2015}; but see \citealt{Bowler2017}). The system, called CR7, is a very bright Ly $\alpha$ emitter at $z=6.6$ with $L_{\mathrm{Ly}\alpha} \sim 10^{44} \mathrm{erg \,s^{-1}}$ \citep{Matthee2015, Sobral2015, Bowler2017}. \citet{Sobral2015} have identified CR7 as a combination of three components: Two clumps which appear to be evolved galaxies in close proximity to a third clump, which provides the vast majority of the Ly $\alpha$ flux. This third clump has been successfully modelled by \citet{Agarwal2016a} as a $M_{\bullet} \sim 4.4 \times 10^6 \, \mathrm{M}_{\odot}$ BH formed through direct collapse around $z\sim20$. Recent work has shown either an active galactic nucleus (AGN) or a young starburst population are also likely explanations for the observed characteristics of CR7 \citep{Bowler2017}. With the former being potentially seeded via the stage of a DCBH, and the latter not requiring a DCBH at all.

The evolution during the initial stages of a potential DCBH-systems such as CR7 is unknown and yet likely consists of a constant interplay between star formation and BH growth. The formation of stars in proto-galaxies is driven by the accretion of gas and subsequent gravitational collapse. The star formation law relating the star formation rate (SFR) surface density in a disc galaxy to its gas surface density \citep{Schmidt1959}, once confirmed by observations of local galaxies \citep{Kennicutt1998}, has more recently been shown to extend to $z\sim1.5$ \citep[see, e.g.][]{Carilli2013}. At higher redshifts, the higher densities imply shorter cooling times (as $t_{\rm cool} \propto \rho^{-1}$). Rapid cooling means that the SFR is only limited by the total gas accretion rate and the growth rate of gravitational instabilities in a galaxy \citep{Dekel2009,Dekel2013}. Previous studies on galaxy evolution and star formation have related the global SFR to disc properties via the growth rate of instabilities \citep{Fall1980, Lacey1983, Wang1994, Schaye2004, Elmegreen2010}. For example, \citet{Elmegreen2010} modeled the growth in gas mass and turbulence driven by gas accretion onto a galaxy and found the SFR was mainly a function of the gas accretion rate. However, star formation is a local process. Star formation can only take place where the gas is unstable to gravitational collapse \citep{Wang1994}.  In this context the BH may also play an important role in the stability of the disc \citep{Lodato2012}. The gas properties will change throughout the galaxy with some regions being more unstable than others. Indeed, \citet{Schaye2004} found that if disc galaxies are rotational supported against collapse this will be particularly at large radii, limiting the radial extent of star formation to within some truncation radius.

A further complicating factor for the growth of proto-galaxies around DCBHs is that the hosting halo is generally in the vicinity of a more massive halo it is likely to merge with at a later stage during its evolution \citep{Agarwal2014a}. During the satellite-phase the provision of fuel for star formation will cease due to stripping processes in the environment \citep{VanDenBosch2008}. The growth of the host galaxy will thus be affected and in turn the path to the locally observed BH-galaxy correlations.

The aim of this paper is two-fold, we want to model the stabilising effect of DCBHs on the gaseous disc in proto-galaxies and their impact on the onset of star formation, and based on these models present arguments for the evolution of DCBHs toward locally observed correlations with host galaxies. First, we lay out the star formation model we use which relates star formation rate to disc instabilities (section~\ref{starform}). The model is first introduced by discussing a non-evolving case in section~\ref{Non-evolving} before being fully explored in section~\ref{Evolving} with evolving the halo and stellar mass. Finally, we discuss the implications of the model for massive seed hosting galaxies and the onset of star formation within them (section~\ref{conclusions}). Throughout the paper a $\Lambda$CDM Universe is assumed with $\rm H_{0} = 70 km\, s^{-1}\, Mpc^{-1}$, $\Omega_{\mathrm{m, \, 0}} = 0.27$ and $\Omega_{\mathrm{\Lambda, \, 0}} = 0.73$.

\section{Star Formation} \label{starform}

The empirical star formation law derived from local observation \citep{Kennicutt1998},

\begin{equation}
    \dot{\Sigma}_{\star}^{\rm KS}(t) = 1.515 \times 10^{-4} \, \mathrm{M_{\odot} \, yr^{-1} \,kpc^{-2}} \left(\Sigma_{\mathrm{g}}/1 \, \mathrm{M_{\odot} \, pc^{-2}}\right)^{1.4}
    \label{eq:sf10}
\end{equation}

\noindent where the amplitude has been adjusted to fit with a \citet{Chabrier2003} IMF \citep{Schaye2010}, has been seen to hold to high redshift \citep[see, e.g.][and references therein]{Carilli2013}. Furthermore, the relation appears to hold for both local surface density values and those integrated over an aperture \citep{Kennicutt1998}. One can understand this relation between star formation rate surface density and gas density using a star formation timescale \citep[see, e.g.][]{Wang1994,Kennicutt1998,Krumholz2007,Elmegreen2010}. The local dynamical or free-fall time within a star forming region is often used to relate the timescale to the gas density while the different mechanisms that would work against gravitational collapse, such as thermal and rotational support, are factored in either explicitly \citep{Wang1994,Elmegreen2010} or as part of an efficiency parameter \citep{Krumholz2007}.

In contrast to global models of star formation in proto-galactic discs \citep[e.g.][]{Elmegreen2010} we are here focusing on the radial star formation profile, which depends on local gravitational instabilities in the disc and allows the investigation of the impact of massive seed BHs.

The Ansatz for the star formation model we use is: (1) Star formation can only take place above a minimum threshold $\Sigma_{\mathrm{g}} > \Sigma_{\mathrm{th}} = 10.0 \, \mathrm{M_{\odot} \, pc^{-2}}$ \citep{Schaye2004}. (2) No star formation will take place if the disc is locally stabilised against gravitational collapse $Q_{\rm Toomre} \geq 1$ or (3)  if the local density is too low to overcome tidal forces $Q_{\rm tidal} \geq 1$ (see section~\ref{modelsetup}).

If these conditions for star formation are met, the SFR surface density is calculated by relating the timescale for star formation to the maximal growth rate of axisymmetric perturbations, $\omega_{\rm WS}$ \citep{Wang1994}:

\begin{equation}
    t_{\rm SF} = \frac{1}{\omega_{\rm WS}} = \frac{Q_{\rm Toomre}}{\kappa \sqrt{1 - Q_{\rm Toomre}^{2}}}
    \label{eq:sf8}
\end{equation}

\noindent where $\kappa$ is the epicyclic frequency and $Q_{\rm Toomre}$ is the Toomre disc instability parameter (see section~\ref{Non-evolving}). The SFR surface density $\dot{\Sigma}_{\star}$ can then be written as a function of this timescale, the gas surface density profile $\Sigma_{\mathrm{g}}$, and the star formation efficiency parameter $\epsilon$, to obtain the following \citep{Wang1994}:

\begin{equation}
\begin{aligned}
    \dot{\Sigma}_{\star}^{\rm WS}(t) = \epsilon \frac{\Sigma_{\mathrm{g}}}{t_{\rm SF}} &= \epsilon \frac{ \kappa}{Q_{\rm Toomre}} \Sigma_{\mathrm{g}} \left(1 - Q_{\rm Toomre}^{2}\right)^{0.5} \\
    &= \epsilon \frac{ \pi G \Sigma_{\mathrm{d}}}{\sigma} \Sigma_{\mathrm{g}} \left(1 - Q_{\rm Toomre}^{2}\right)^{0.5}
    \label{eq:sf9}
\end{aligned}
\end{equation}

\noindent where $\sigma$ is the velocity dispersion (from here on taken to be the sound speed, $c_s$) and the $\Sigma_{\rm d} =\Sigma_{\mathrm{g}} + \Sigma_{\star}$ is the total surface density of the disc. This formulation includes explicitly the effects of rotation on the growth rate of instabilities in the disc. The rotation of the disc provides support against gravitational instabilities, preventing the collapse of gas to form stars \citep{Fall1980, Lacey1983}. This then relates the SFR to the growth rate of gravitational instabilities rather than simply the free-fall timescale and allows one to take the structure of the disc in to account. \footnote{Equation~\ref{eq:sf9} will result in a steeper $\dot{\Sigma}_{\star} - \Sigma_{\mathrm{g}}$ relation ($\dot{\Sigma}_{\star} \propto \Sigma_{\mathrm{g}}^{2}$) than the \citet{Kennicutt1998} law (equation~\ref{eq:sf10}) and is therefore likely to overestimate the total SFR, providing a conservative, upper-limit estimate for our model which we are seeking for in this study.}

\section{Non-evolving, non-star forming case} \label{Non-evolving}

\subsection{Model Setup} \label{modelsetup}

This first case looks at the effect of the BH in a non-evolving gaseous disc (The fully evolving case is addressed in section~\ref{Evolving}). The mass of the halo is kept constant, $\dot{M}_{\rm 200} = 0$, and the stellar mass is zero throughout, $M_{\star}=0$.
To investigate the effects of a seed BH on a galaxy forming in its host halo it is necessary to probe the inner region of the proto-galactic disc. For this reason we take the radial dependencies of system's properties into account. For the non-evolving case we have a purely gaseous disc embedded in a halo we model simply as an isothermal sphere\footnote{Whether the inner density profile of a dark matter halo in this scenario should be less steep is not clear due to the effects of baryons on the halo \citep[see, e.g.][]{Davis2014}.} (see Section~\ref{stability} for a discussion of the implications of this). We assume the gas disc has an exponential profile centred on the BH such that the surface density of gas goes with radius as,

\begin{equation}
	\Sigma_{\mathrm{g}}(R) = \Sigma_{g,\,0} \exp(-R/R_{\mathrm{d}})
	\label{eq:sigprofile}
\end{equation}

\noindent $R_{\rm d}$ is the disc scale radius which is set by the halo parameters and $\Sigma_{g,\,0} = M_{\rm g} / (2 \pi R_{\rm d}^2 )$. As we assume the halo is an isothermal sphere, the disc scale radius is calculated using the following \citep{Mo1998}:

\begin{equation}
	R_{\mathrm{d}} = \frac{1}{\sqrt{2}} \left (\frac{j_{\rm d}}{m_{\rm d}} \right) \,\lambda\, r_{\rm 200}
	\label{eq:Rd}
\end{equation}

\noindent Throughout this study $j_{\rm d}/m_{\rm d}$ is assumed to be unity and the spin parameter is taken from the log-normal distribution used by \citet{Mo1998} with the first and second moments: $\bar{\lambda}=0.05$ and $\sigma_{\lambda}$ respectively. For simplicity the first moment of the distribution is used as our fiducial value for $\lambda$ unless otherwise stated, though it is important to note changing $\lambda$ will have an effect on the model. For example, taking $\lambda$ at the 10\% point of the distribution roughly halves the disc scale radius which, for the same disc mass, doubles the surface density, $\Sigma_{g,\,0}$. For further discussion in the context of the model see Appendix~\ref{app:rd}.

The local stability of the disc against gravitational collapse is parametrised by the Toomre stability parameter \citep{Toomre1964}. If we have a disc which has both a stellar and gas component and assume the velocity dispersion of each component is such that $\sigma \equiv \sigma_{\mathrm{g}} = \sigma_{\mathrm{s}}$, the Toomre parameter becomes:

\begin{equation}
	Q_{\rm Toomre} = \frac{\kappa\,\sigma}{\pi\,G\,\Sigma_{\mathrm{d}}}
	\label{eq:toomre}{}
\end{equation}

\noindent where $\kappa$ is the epicyclic frequency and $\sigma$ is the velocity dispersion \citep{Toomre1964, Wang1994, Romeo2011}. If $Q_{\rm Toomre}>1$ the disc is stable to gravitational instabilities; $Q_{\rm Toomre}<1$ the disc is unstable and $Q_{\rm Toomre}\sim1$ the disc is partially stable \citep[see, e.g.][]{Lodato2007}. The velocity dispersion is taken as the sound speed of the gas, $c_{\rm s}$. This acts as a lower limit as we do not take turbulent motions into account, however at $T_{\rm g}= 8000$ K, the sound speed should provide a significant fraction of the gas velocity dispersion\footnote{With $\sigma = c_{\rm s}$, Equation~\ref{eq:toomre} is similar to $Q_{\rm Thermal}$ as described recently by \citet{Stark2018}.}. The epicyclic frequency describes the rotational support of the system due to the gravitational potential. It can be expressed as a function of the angular velocity, $\Omega = V_{\rm c} / R$ (where $V_c$ is the circular velocity which is calculated from the radial derivative of the gravitational potential):

\begin{equation}
	\kappa^{2} = \frac{2 \,\Omega}{R}\frac{d}{dR}(R^{2}\Omega)
	\label{eq:kappa}
\end{equation}

\noindent This is therefore a function of the three components of the system: the halo, disc and BH. As the potential due to these components can be combined to find the full potential, $\kappa$ can be split into the corresponding parts. The relative importance of the component of $\kappa$ due to the BH will increase with proximity to the BH.

Similarly to the Toomre parameter, the critical tidal density, $\rho_{\mathrm{tidal}}$, defines the limit to the local density of the disc below which the local self gravity of the disc is weak compared to the tidal forces on the disc \citep{Hunter2001,Martig2009}.

\begin{equation}
	\rho_{\mathrm{tidal}}=\frac{3 \, \Omega \, R}{2 \pi G}\abs*{\frac{d\Omega}{dR}}
	\label{eq:rhotidal}
\end{equation}

 Dominant tidal forces inhibit the growth of density perturbations, preventing stars from forming. To compare the tidal and Toomre stability of the disc we need to make a comparison between the critical tidal density and the Toomre parameter. We define the critical tidal surface density as $\Sigma_{\mathrm{tidal}} = 2 H \rho_{\mathrm{tidal}}$ and using the scale height of the disc, $H = \sigma^2/(\pi\,G\,\Sigma_{\mathrm{d}})$, we obtain:

\begin{equation}
    \Sigma_{\mathrm{tidal}}=3 \Omega R \frac{\sigma^2}{\pi^2 G^2 \Sigma_{\mathrm{d}}}\abs*{\frac{d\Omega}{dR}}.
    \label{eq:sdtidal}
\end{equation}

\noindent With some rearranging we can see the square root of the ratio of the critical tidal surface density to the local surface density is of a similar form to the Toomre parameter.

\begin{equation}
    Q_{\rm tidal} \equiv \sqrt{\frac{\Sigma_{\mathrm{tidal}}}{\Sigma_d}}=\frac{\sigma}{\pi G \Sigma_{\mathrm{d}}}\sqrt{3 \Omega R\abs*{\frac{d\Omega}{dR}}}. = Q_{\rm Toomre} \frac{\nu}{\kappa} 
    \label{eq:sdratio}
\end{equation}
\noindent where we define the tidal frequency $\nu$,
\begin{equation}
    \nu^2 = 3 \Omega R\abs*{\frac{d\Omega}{dR}}
    \label{eq:tidalfreq}
\end{equation}

\noindent It follows that $Q_{\rm tidal}$ will behave similarly to the Toomre parameter; if $Q_{\rm tidal} < 1$ it implies $\rho > \rho_{\mathrm{tidal}}$ and the disc's local self gravity dominates but if $Q_{\rm tidal} > 1$, $\rho < \rho_{\mathrm{tidal}}$ and the disc is locally unstable to tidal forces, and the growth of gravitational instabilities locally in the disc is inhibited.
We combine the two stability parameters by defining $Q_*$ as the maximum of the two:

\begin{equation}
    Q_*=\mathrm{max}[Q_{\rm tidal},\;Q_{\rm Toomre}] 
    \label{eq:Qmax}
\end{equation}
\noindent this reduces our conditions for star formation down to two: that the surface density is above the threshold ($\Sigma_{\mathrm{g}} > \Sigma_{\mathrm{th}} = 10.0 \, \mathrm{M_{\odot} \, pc^{-2}}$) and that $Q_*>1$.

\begin{table}
	\centering
	\caption{Table of non-evolving, non-star forming model parameters.}
	\label{tab:nonevopara}
	\begin{tabular}[h]{l|l|l} 
		\hline
		Parameter & Definition & Fiducial\\
		\hline
		$M_{200}$ & dark matter halo mass ($\mathrm{M_{\odot}}$) & $5\times10^{8}$ \\
		$\lambda$ & halo spin parameter& 0.05\\
		$j_{\rm d}/m_{\rm d}$ & disc and halo specific momenta ratio& 1.0\\
		$f_{\rm b}$ & baryon fraction & 0.17\\
		$T_{\rm g}$ & gas temperature (K)& 8000\\
		$z$ & redshift& 10.0\\
	\end{tabular}
\end{table}

\subsection{Stability Parameters Profiles} \label{stability}

For our analysis here we use a fiducial model of an atomic cooling halo at $z\sim10$ (see Table~\ref{tab:nonevopara}). The total mass of the system was calculated for an atomic cooling halo $M_{\rm total}\sim M(T_{\rm vir}=8000\,K)$ \citep{Mo2010} and the disc mass was calculated by taking the baryonic mass of the halo, $M_{\rm d} = f_{\rm b} M_{\rm total}$ where $f_{\rm b}=0.17$ is the universal baryon fraction, unless otherwise stated. One would expect only a fraction of the baryonic mass of the halo, $p\,f_{\rm b}$ (where $p<1$), to reach the disc \citep[see, e.g.][]{Dekel2013a, Dekel2013} and therefore, $f_{\rm b} M_{\rm total}$ is an upper limit on the disc mass. Taking this upper limit allows us to look at the most unstable case as lowering the disc mass would increase the disc stability. For our model, decreasing $p$ would have the same effect as decreasing the overall baryon fraction. Section~\ref{minQ} discusses how lowering the baryon fraction would change the stability profile of the disc for different BH and disc masses.

\begin{figure}
	\includegraphics[width=\columnwidth]{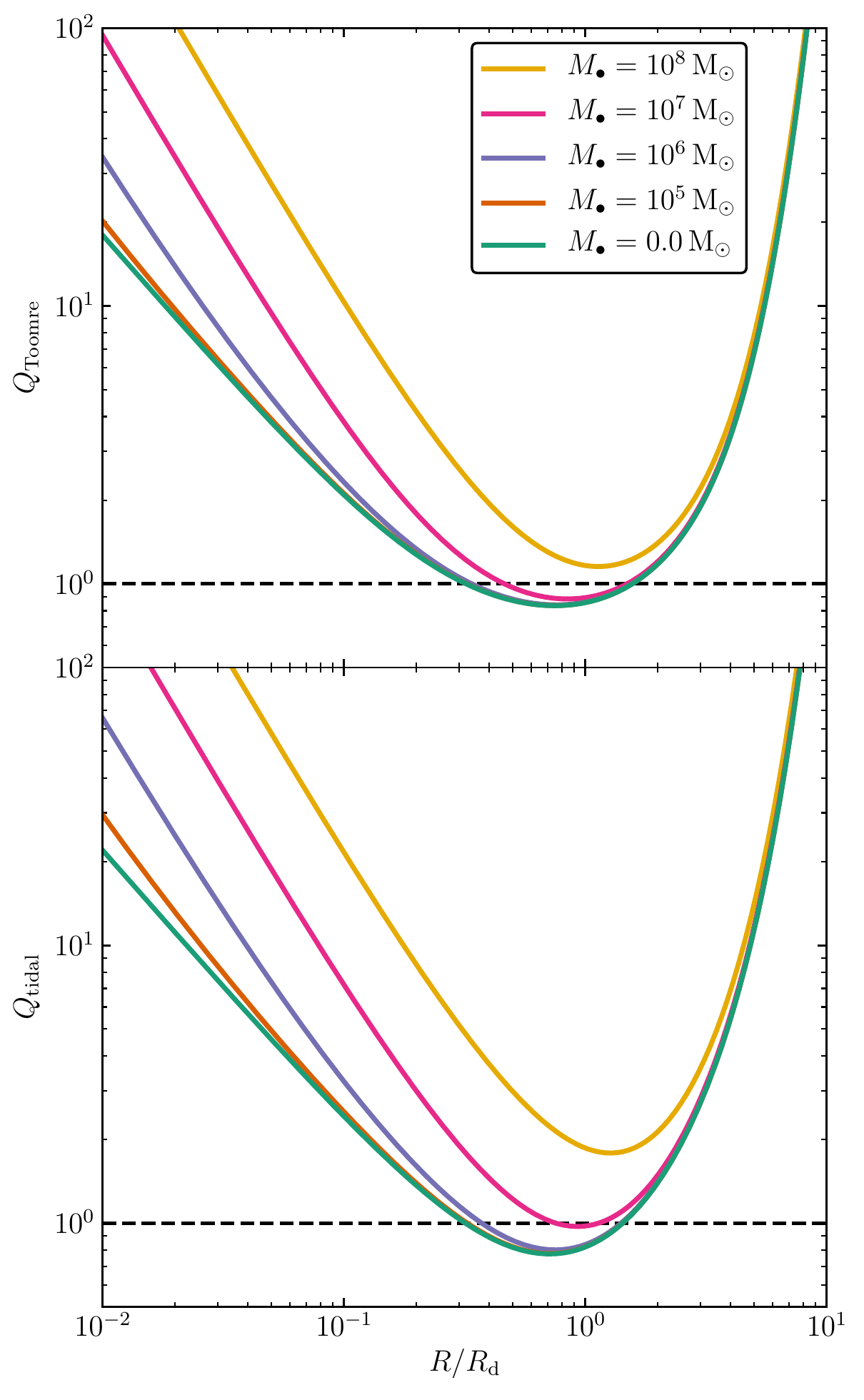}
    \caption{The stability parameters profiles for the same disc without a BH and for four different mass BHs. The top panel shows the Toomre stability parameter profile and the bottom shows the tidal stability parameter profile. The disc is the same in all cases with $M_{\rm d} = 1.02\times10^8 \, \mathrm{M_{\odot}}$ and $R_{\rm d} = 86.5 \, \mathrm{pc}$.}
    \label{fig:Qprofile}
\end{figure}

Figure~\ref{fig:Qprofile} shows the radial profiles for the Toomre and tidal stability parameters respectively for our fiducial set-up as summarized in table \ref{tab:nonevopara}. If we first look at the upper panel of Figure~\ref{fig:Qprofile}, the increase in $\kappa$ at small radii due to the presence of the BH stabilises the inner-most region of the disc, shown by the increase in $Q_{\rm Toomre}$ at small radii. Increasing the BH mass increases this effect, narrowing the region of the disc where star formation can take place. For a constant disc mass an accreting BH thus would be able to prevent a larger fraction of the disc from forming stars as it grows in mass. At larger radii the influence of the BH diminishes and the disc determines the shape of the stability profiles except for cases with the largest BH masses. After reaching a minimum both of the stability parameters increase as the disc surface density decreases with radius.

The tidal stability parameter profile (the lower panel of Figure~\ref{fig:Qprofile}) shows how the BH also has a strong tidal effect on the disc at small radii. The tidal and Toomre parameter profiles are similar in shape however, the tidal parameter appears to be below the critical value of 1 over a narrower range in radius. This suggests $Q_{\mathrm{tidal}} \leq 1$ is a stricter condition for star formation in the disc within the model than simply $Q_{\rm Toomre} \leq 1$. 

\begin{figure}
	\includegraphics[width=\columnwidth]{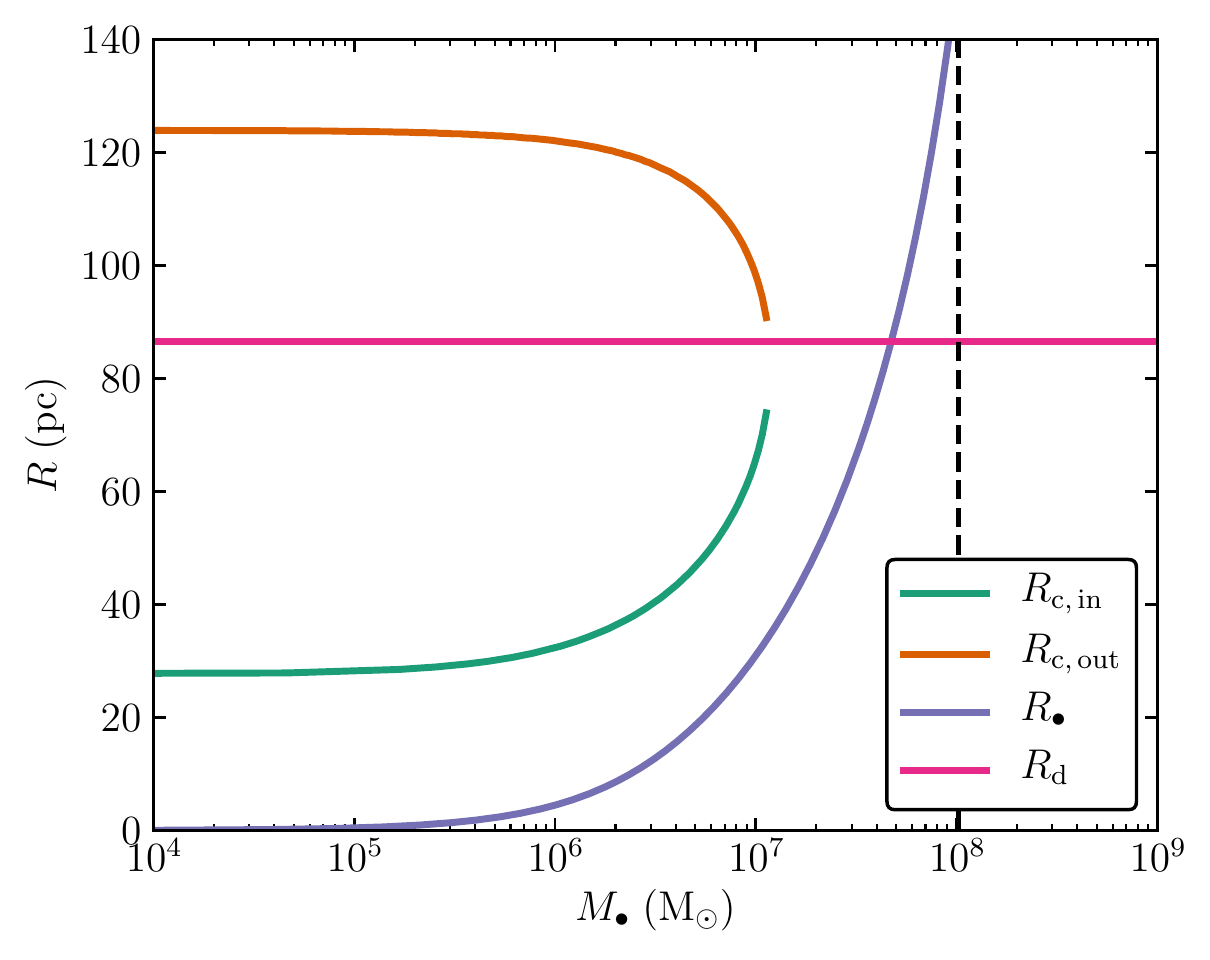}
    \caption{The relationship between characteristic radii versus BH mass for a disc with $M_{\rm d} = 1.02\times10^8 \, \mathrm{M_{\odot}}$ (shown as the vertical, dashed line) and a scale radius of $R_{\rm d} = 86.5 \, \mathrm{pc}$. The disc scale radius is shown in pink, the sphere of influence radius of the BH ($R_{\rm \bullet}$) is shown in purple, the inner and outer critical radii ($R_{\rm c, \, in}$ and $R_{\rm c, \, out}$) are shown in green and orange respectively. The green and orange lines stop at $M_{\bullet} \sim 10^7 \, M_{\odot}$, the mass at which the BH completely stabilises the disc.}
    \label{fig:RvsMbh}
\end{figure}

For a constant disc mass the inner critical radius, the inner-most radius where star formation can occur (where both $Q_{\rm Toomre}$ and $Q_{\rm tidal}$ are $\leq 1$), increases with BH mass (Figure~\ref{fig:RvsMbh}). The inner critical radius, $R_{\rm c, \, in}$, is smallest when we have no BH at $R_{\rm c, \, in} = 27.3 \, \mathrm{pc}$ and is $R_{\rm c, \, in} = 32.3 \, \mathrm{pc}$ for $M_{\bullet} = 10^6 \, \mathrm{M_{\odot}}$. If we increase the BH mass enough we reach a point where the whole disc becomes stabilised (For example, the yellow $M_{\bullet} = 10^8 \, \mathrm{M_{\odot}}$ BH case in Figure~\ref{fig:Qprofile}). The point where the disc becomes completely stabilised is represented in Figure~\ref{fig:RvsMbh} by the point where the $R_{\rm c, \, in}$ lines end around $M_{\bullet} \sim 10^7 \, \mathrm{M_{\odot}}$ which is less than a tenth of the disc mass ($M_{\rm d} = 1.02\times10^8 \, \mathrm{M_{\odot}}$). 
It can be useful to compare this inner critical radius to the radius of the sphere of influence of the BH, $R_{\bullet}$. We calculate this from $M_{\bullet}$ and the circular velocity profile of the system, $V_c(R)$, using:
\begin{equation}
    R_{\bullet} = \frac{G M_{\bullet}}{V_c^2}
    \label{eq:rbullet}
\end{equation}
\noindent Due to the radial dependence of $V_c(R)$ it is necessary to solve Equation~\ref{eq:rbullet} iteratively such that $V_c=V_c(R_{\bullet})$.

As we increase the BH mass towards this disc-stabilising value, $R_{\rm c, \, in}$ and $R_{\rm c, \, out}$ are of the same order as $R_{\rm d}$, with $R_{\rm c, \, in}$ roughly 2.5 times $R_{\bullet}$.

The radius at which either stability parameter is minimized is always of the order of the disc scale radius (see Figure~\ref{fig:Rqmin} and the following section). Starting at BH mass fractions of $10\%$ the radius at which the tidal stability parameter is minimized quickly catches up with the scale radius of the disc. This helps to explain why $R_{\rm c, \, in}$ approaches $R_{\rm d}$ as the disc becomes stabilised as we increase the BH mass.

As the disc approaches stability the minimum value of $Q_{\rm tidal}$ approaches 1 until the BH mass is sufficiently massive to fully stabilise the disc ($M_{\bullet}/M_{\rm d}\sim 10\%$) and $R_{\rm c, \, in} = R_{\rm c, \, out} = R_{\rm Q_{min}}$. At lower BH masses, the inner critical radius is generally less than the disc scale radius but it can be significantly greater than $R_{\bullet}$ depending on the masses of the BH and disc.

It is important to note that the halo profile also plays a role in determining the shape of the $Q$ profiles. In the case of a less centrally dominated halo profile, such as the NFW profile \citep{Navarro1996}, the inner disc would be more unstable in the absence of a massive BH, decreasing $R_{\rm c, \, in}$. However, the BH would provide a more signifcant fraction of the total mass in the inner region of the system and the relative stabilising effect of the BH would therefore be enhanced in a NFW profile halo.

\subsection{Minima of the Stability Parameter Profiles} \label{minQ}

\begin{figure}
	\includegraphics[width=\columnwidth]{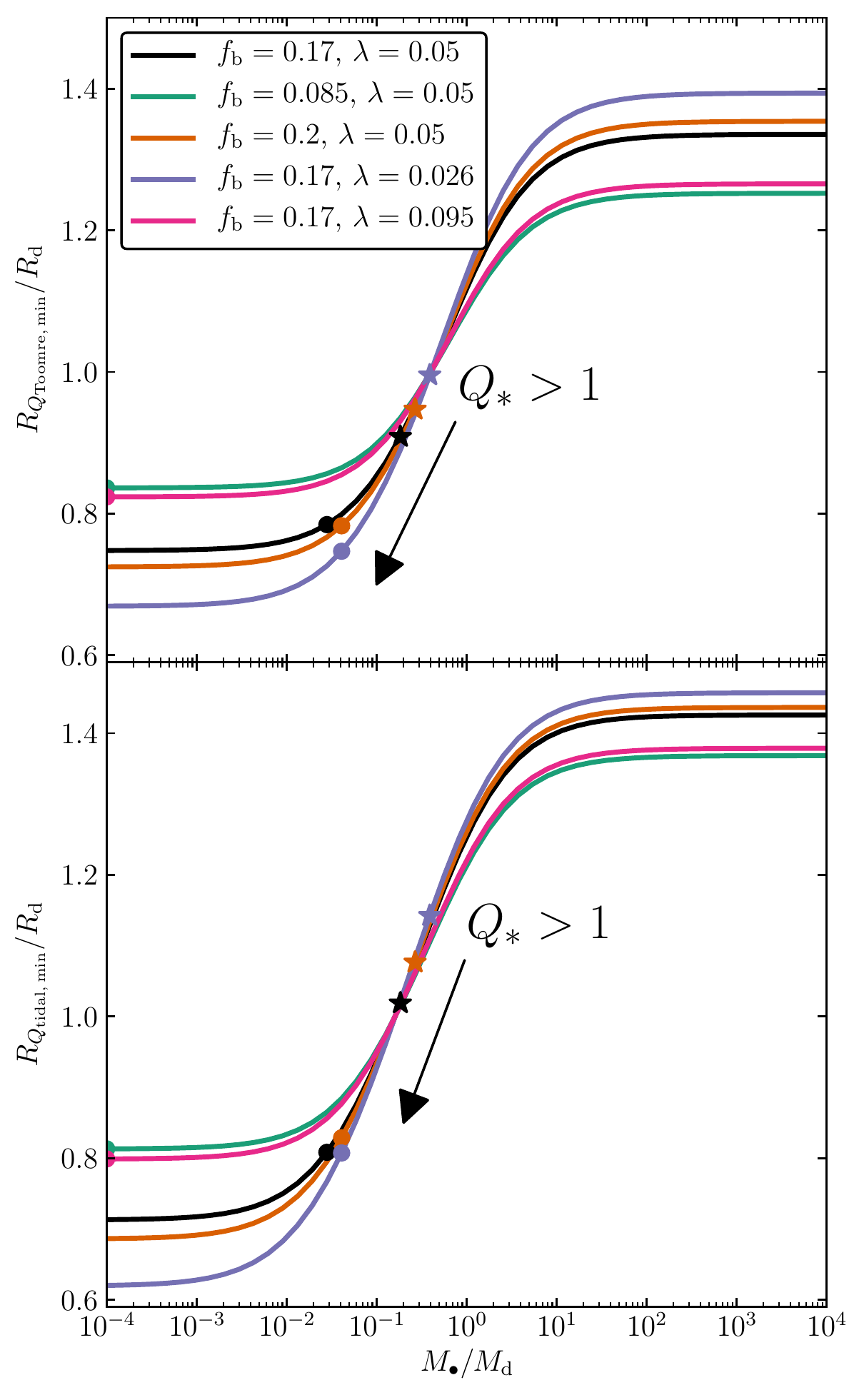}
    \caption{The top panel shows radius where the Toomre stability parameter is minimized as a fraction of the disc scale radius as a function of the mass ratio of the BH and disc. The bottom panel shows the same for the tidal stability parameter. The black line was found for the fiducial case of $f_{\rm b} = 0.17$, $ \lambda=\bar{\lambda}\equiv0.05$. The remaining lines represent cases where either $f_{\rm b}$ or $\lambda$ are changed from fiducial case to the values indicated in the legend. Unlike the curves, the position of the markers are dependent on the total mass of the system. The $M_{\rm total}=10^9 \mathrm{M}_{\odot}$ case is shown as an example. The dots correspond to the point where $Q_{\rm tidal,\, min}=Q_{\rm Toomre, \, min}$ i.e. to the left of these dots the Toomre parameter is a stricter criterion for star formation and to the right the tidal parameter is more the strict of the two. The stars correspond to the point where $Q_{*,\,\rm min}=1$. There are no star symbols shown for the low baryon fraction and high spin parameter cases (green and pink lines respectively) as the disc is fully stable for both, even in the absence of a BH.}
    \label{fig:Rqmin}
\end{figure}

The radii of the minimum of the two stability parameter profiles as a fraction of the disc scale radius are only a function of the BH-to-disc mass ratio, the baryon fraction and the spin parameter of the halo (Assuming $j_{\rm d}/m_{\rm d} = 1$). Each of the lines in Figure~\ref{fig:Rqmin} represent a different combination of baryon fraction and spin parameter. While changing the total mass of the system will change the absolute value of $Q_*$, it does not change the positions (as a fraction of $R_{\rm d}$) of the minimum values of $Q_{\rm Toomre}$ and $Q_{\rm tidal}$. That is the ratios $f_{\rm b}$ and $M_{\bullet}/M_{\rm d}$ define the relative importance of the different components of each $Q$ and therefore shape of the $Q$ profiles. The spin parameter $\lambda$ defines $R_{\rm d}$ and therefore changes the surface density of the disc. Hence, the $R_{Q_{\rm min}}/R_{\rm d}$ -- $M_{\bullet}/M_{\rm d}$ relationships are influenced by $\lambda$ as the disc surface density and velocity profile (and therefore the disc stability) are dependent on it. Increasing the baryon fraction or decreasing the spin parameter leads to a similar change in the $R_{Q_{\rm min}}/R_{\rm d}$ -- $M_{\bullet}/M_{\rm d}$ curves. The curves shown with the higher and lower spin parameters correspond to the upper and lower limits of the 80\% confidence interval of the $\lambda$ probability distribution \citep{Mo1998}. The upper limit to the baryon fraction is unlikely to be much greater than our fiducial value \citep{Qin2017}. Therefore, $f_{\rm b}=0.2$ would be an extreme case. Though a lower baryon fraction than $f_{\rm b}=0.085$ is possible \citep{Qin2017}, such a system would struggle to have an unstable disc in our model. Over this range of values, the lower spin parameter limit case,  $\lambda = 0.026344$, has the largest range in $R_{Q_{\rm min}}/R_{\rm d}$, with a factor of $<2.5$ change. This indicates $R_{Q_{\rm min}}\sim R_{\rm d}$ over the relevant parameter space.

The curves in Figure~\ref{fig:Rqmin} were found to follow the functional form:

\begin{equation}
\frac{R_{Q_{\rm min}}}{R_{\rm d}} = A-B \tan ^{-1}\left[C \left(\frac{M_{\bullet}}{M_{\rm d}}\right)^{-D}\right]
\end{equation}\label{eq:rqminfunc}

See Table~\ref{tab:fitdata} for the values corresponding to each curve in Figure~\ref{fig:Rqmin}.

For all curves, $R_{Q_{\rm tidal,\, min}}/R_{\rm d}$ -- $M_{\bullet}/M_{\rm d}$ is steepest between $M_{\bullet}/M_{\rm d}\sim0.1$ and $M_{\bullet}/M_{\rm d}\sim2$ and the points found for $Q_{*,\,\rm min}=1$ all lie in that range. Note, for $M_{\rm total}=10^9\, \mathrm{M}_{\odot}$, the tidal parameter becomes the more strict criterion at $M_{\bullet}/M_{\rm d} < 0.1$ in each case shown.
\begin{table}
	\centering
	\caption{Table of the $R_{Q_{\rm min}}/R_{\rm d}$ -- $M_{\bullet}/M_{\rm d}$ fit parameters for each curve in Figure~\ref{fig:Rqmin}. The fiducial case is shown in the top row.}
	\label{tab:fitdata}
	\begin{tabular}[h]{c|c|c|c|c|c}
		\hline
        & &
        \multicolumn{4}{c}{$Q_{\rm Toomre}$} \\
		$f_{\rm b}$ & $\lambda$ & A & B & C & D\\
		\hline
        0.17 & 0.05 & 1.43 & 0.4596 & 0.3904 & 0.6898 \\
        \hline
        0.085 & 0.05 & 1.371 & 0.3571 & 0.4572 & 0.7338\\ 
        0.2 & 0.05 & 1.441 & 0.4842 & 0.3773 & 0.6782\\
        0.17 & 0.026344 & 1.463 & 0.5414 & 0.3505 & 0.6496 \\
        0.17 & 0.094898 & 1.381 & 0.3731 & 0.4454 & 0.7275 \\
        \hline
        & & \multicolumn{4}{c}{$Q_{\rm tidal}$} \\
		$f_{\rm b}$ & $\lambda$ & A & B & C & D\\
        \hline
        0.17 & 0.05 & 1.338 & 0.3776 & 0.6429 & 0.7381\\
        \hline
        0.085 & 0.05 & 1.254 & 0.2669 & 0.7043 & 0.7739\\
        0.2 & 0.05 & 1.357 & 0.4046 & 0.6302 & 0.7281\\
        0.17 & 0.026344 & 1.398 & 0.4665 & 0.6028 & 0.7031\\
        0.17 & 0.094898 & 1.267 & 0.2836 & 0.6938 & 0.7691 \\
	\end{tabular}
\end{table}

\begin{figure*}
	\centering
	\includegraphics[width=\textwidth]{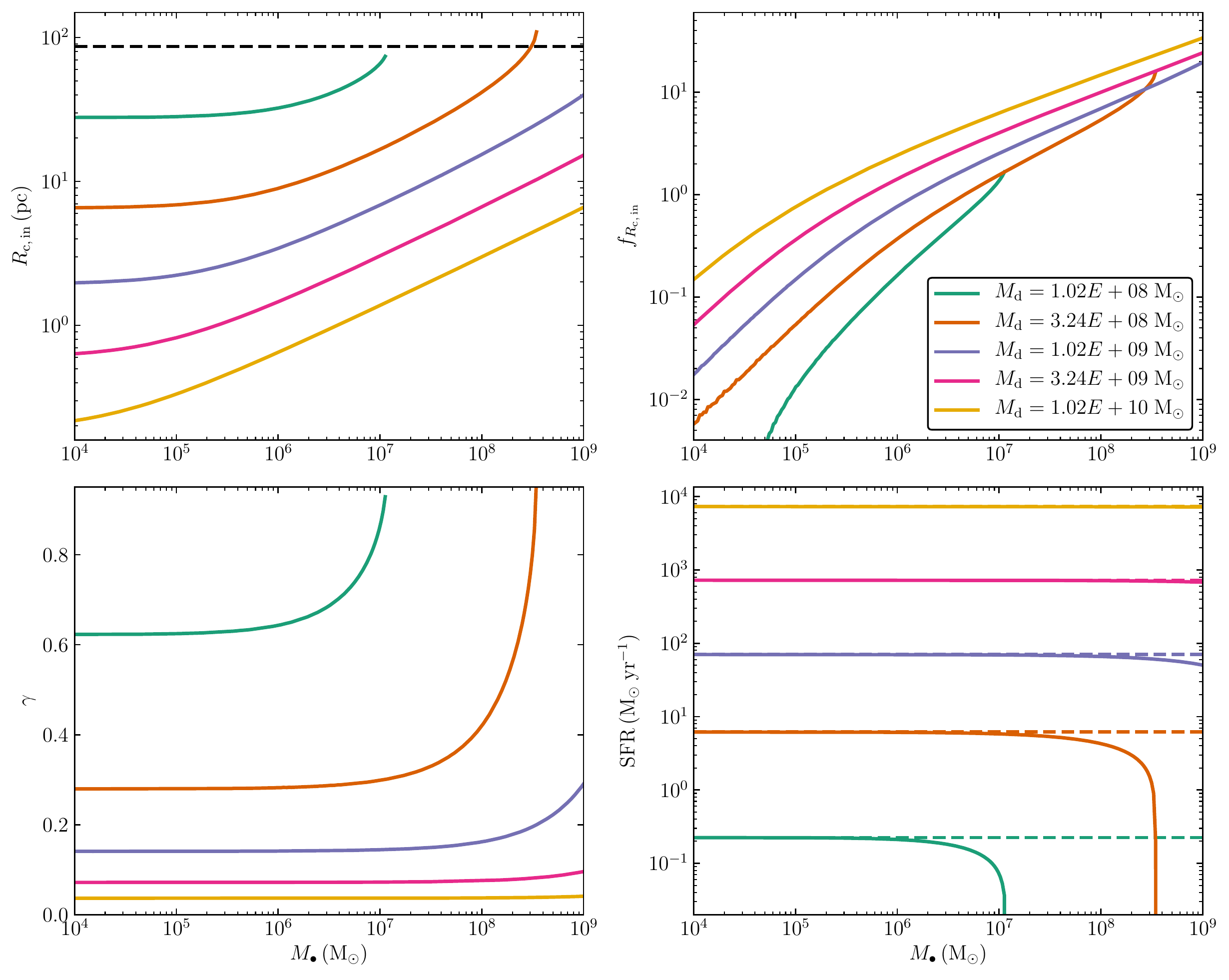}
    \caption{Each of the panels show a property of a disc relating to its stability as a function of BH mass for different disc masses. The halo mass, $M_{200}=5 \times 10^8\,\mathrm{M}_{\odot}$, and the disc scale radius, $R_{\rm d}=86.5\,\mathrm{pc}$, are the same in all cases. The top left panel shows the inner critical radius. The curves tend to the no BH case at the low mass BH end. The disc scale radius is shown as a black dashed line. The top right panel shows the relative difference in the inner critical radius with and without a BH $f_{R_{\rm c, \; in}}$ (see Equation~\ref{eq:reldiff}). The bottom left panel shows the stable fraction of the disc $\gamma$ (see Equation~\ref{eq:gamma}). The bottom right panel shows the star formation rate (SFR) for both with and without a BH (the solid and dashed lines respectively).}
    \label{fig:RcvsMbh}
\end{figure*}

\subsection{Change in Disc Stability with Black Hole and Disc Mass}

The inner critical radius is shown as a function of BH mass for different disc masses in the top left panel of Figure~\ref{fig:RcvsMbh}. The purple line represents the same disc mass as used in Figures~\ref{fig:RvsMbh}. Increasing the disc mass decreases the critical radius as the disc becomes more unstable. For each case in the top left panel of Figure~\ref{fig:RvsMbh} the curve tends to the no BH case at the low mass BH end. As said above, as the BH mass increases the critical radius does increase, however, this increase behaves slightly differently for the higher disc masses. 
Since the higher disc mass decreases $R_{\rm c, \, in}$, the critical radius is closer to the BH. This in turn means small changes to the BH mass at the low mass end has a greater influence on the value of $R_{\rm c, \, in}$; the relative difference in $R_{\rm c, \, in}$ between the cases with and without a BH is defined as:
\begin{equation}
\begin{aligned}
    f_{R_{\rm c, \, in}}\equiv(R_{\rm c, \, in} - R_{\rm c, \, in}(M_{\bullet}=0))/R_{\rm c, \, in}(M_{\bullet}=0)
    \label{eq:reldiff}
\end{aligned}
\end{equation}
This is shown to increase with disc mass in the top right panel of Figure~\ref{fig:RcvsMbh} which shows the relative difference as a function of BH mass for different disc masses. At higher BH masses the lower mass disc curves are steeper as the BH mass is increasingly comparable with the disc until the disc is fully stabilised. This is not seen in the higher disc mass cases as the BH masses investigated do not reach the range required to stabilise these discs. Note the BH mass required to fully stabilise the $M_{\rm d} = 3.24\times10^8 \, \mathrm{M_{\odot}}$ disc (orange curve) is $M_{\bullet} \gtrsim M_{\rm d}$ whereas the lowest disc mass case needs only $M_{\bullet} \lesssim 0.1 \, M_{\rm d}$. These numbers are in line with the range where the dependence of $R_{Q_{\rm tidal,\, min}}/R_{\rm d}$ on $M_{\bullet}/M_{\rm d}$ is strongest ($0.1 \lesssim M_{\bullet}/M_{\rm d} \lesssim 2$; see previous section).

We define the stable fraction of the disc as fraction of the disc mass outside the unstable region between $R_{\rm c, \; in}$ and $R_{\rm c, \; out}$,

\begin{equation}
\begin{aligned}
    \gamma &\equiv 1 - \frac{M_{\rm g}(<R_{\rm c, \; out}) - M_{\rm g}(<R_{\rm c, \; in})}{M_{\rm g}}
    \label{eq:gamma}
\end{aligned}
\end{equation}

\noindent The bottom left panel of Figure~\ref{fig:RcvsMbh} shows the stable fraction of the disc mass $\gamma$ as a function of the BH mass for the same cases as the other panels in the figure. In the model this is the fraction of the disc that is stabilised against gravitational collapse, i.e. the fraction of the disc that is outside the region where star formation can take place. That the most massive disc case is almost completely unstable is to be expected and the BH has no effect on the stability fraction for this case. Such a disc could not form as in this case as it greatly outweighs its host halo. Looking at the two lowest mass cases, there is a sharp change in the stable fraction of the disc as the BH mass becomes more comparable with the disc mass ($\sim10\%$), in line with trend seen in $R_{\rm c, \, in}$ panel as the disc reaches stability and $R_{\rm c, \, in} \sim R_{\rm d}$.

The bottom right panel of Figure~\ref{fig:RcvsMbh} shows the how the total SFR changes as a function of BH mass for different disc masses. The two cases with the lowest disc masses are where the difference between the cases with and without a black can be most significant. In these cases, the steep drop in SFR we see as the BH mass increases appears to simply reflect the stable gas fraction increase on the adjacent panel. In the lowest disc mass the drop-off occurs with $M_{\bullet} \lesssim 0.1\,M_{\rm d}$ (green line) while with the next higher mass disc the drop-off is at $M_{\bullet}\sim M_{\rm d}$, following the trend in stable fraction.

\subsection{Star Formation Timescale Profile}

\begin{figure}
	\includegraphics[width=\columnwidth]{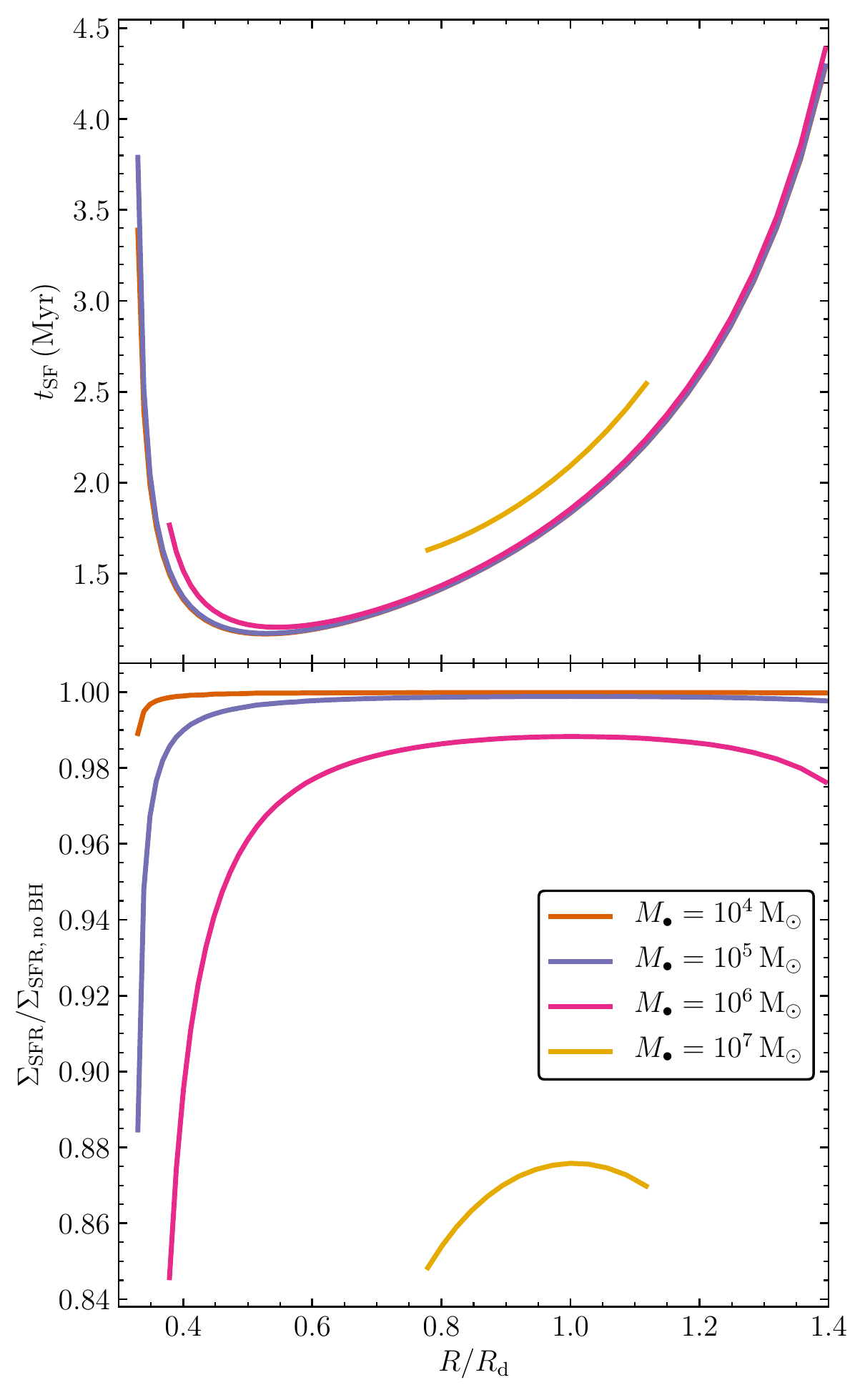}
    \caption{The two panels show the following for a disc of mass $M_{\rm d} = 1.02\times10^8 \, \mathrm{M_{\odot}}$ for different BH masses: The top panel shows the star formation timescale versus radius for different BH masses; the bottom panel shows the ratio of the star formation surface density profiles for with and without a BH for different BH masses.}
    \label{fig:tsfvsx}
\end{figure}

To see how star formation is affected by the mass of the BH in more detail we need to look at the star formation timescale. Not only is the region where star formation can take place constrained by the BH but also the star formation timescale in the model can, in principle, be affected by the presence of the BH. This is because the timescale, $t_{\rm SF}$, is dependent on $Q_{\rm Toomre}$ and therefore $\kappa$ (see Equation~\ref{eq:sf8}), which depends on the BH mass as outlined above. The top panel of Figure~\ref{fig:tsfvsx} shows how the star formation timescale varies as a function of radius for different BH masses and a constant disc mass ($M_{\rm d} = 1.02\times10^8 \, \mathrm{M_{\odot}}$). At a given radius close to the BH, increasing the BH mass increases the timescale until the disc becomes locally stable ($Q_*(R) \geq 1$). There is little variation between the profiles except for the largest BH mass case with $M_{\bullet} = 10^7 \, \mathrm{M_{\odot}}$, where the BH mass is 10\% of the disc mass. Between this case and the lowest BH mass case, the width and area of the star forming region of the disc are around 2/3 smaller while the value of the timescale increases at a given radius by $\sim$15\%. The reduction in the fraction of the gas capable of forming stars at higher BH masses provides a more significant decrease in the star formation rate than the increase in the value of star formation timescale.

A comparison is made with the no BH case in the bottom panel of Figure~\ref{fig:tsfvsx} through finding the ratio of star formation surface density profiles for with and without a BH. The star formation rate in the presence of a BH is less than the no-BH case at all radii and is $\sim$15\% lower close to the BH at $R_{\rm c, \, in}$ for the two most massive BH cases. The presences of the BH changes the SFR surface density profile, and therefore the total SFR, due to the change in the Toomre parameter profile. By limiting the range in radius where stars can form the BH limits the total SFR and would confine the stellar mass to a narrow ring in the disc, ignoring any following redistribution of stars (e.g. through stellar or tidal interactions).

\section{Evolving Halo Model} \label{Evolving}

\subsection{DCBH Hosting Haloes} \label{DCBH haloes}

\begin{figure}
	\includegraphics[width=\columnwidth]{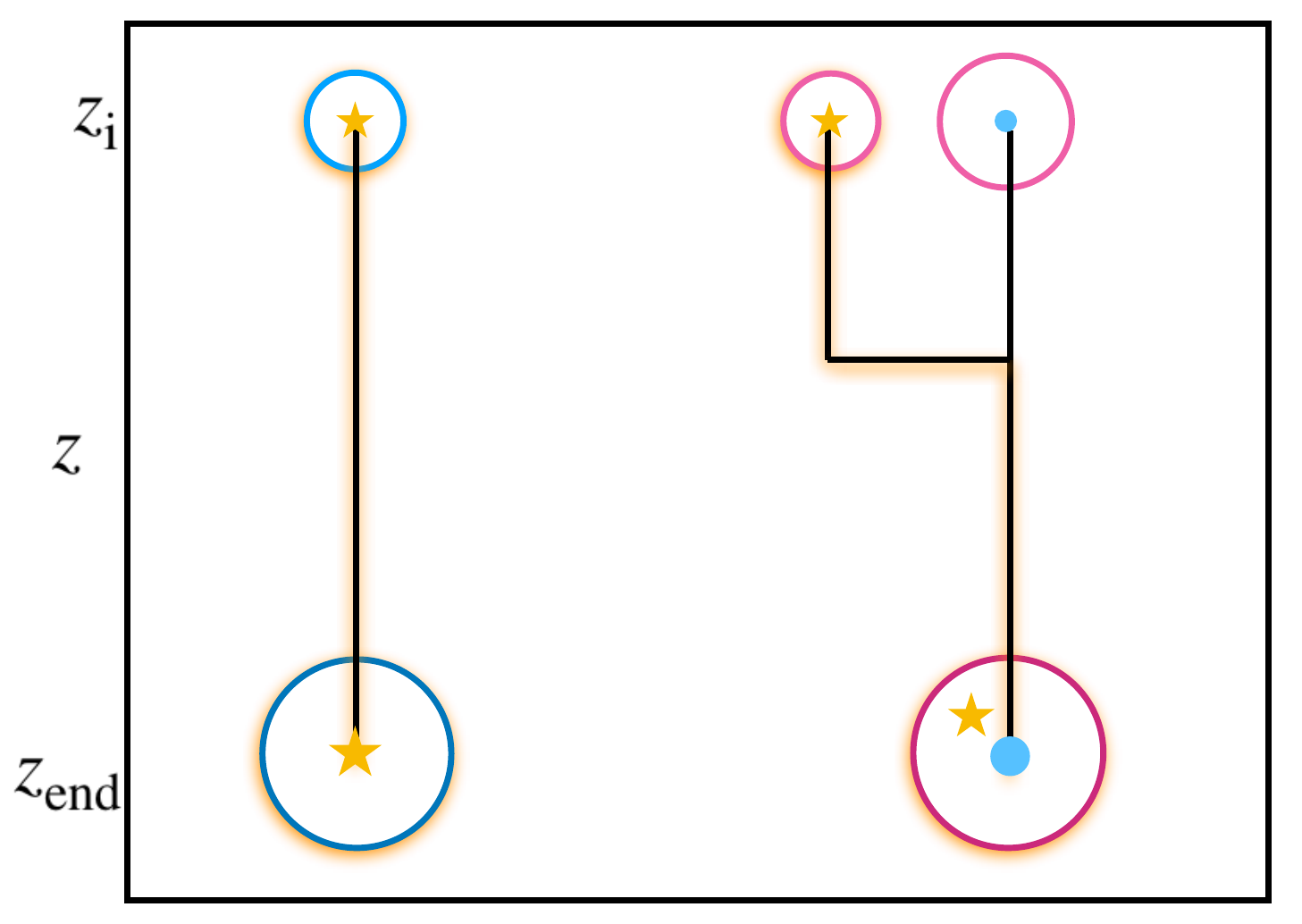}
   \caption{A schematic diagram illustrating the cases investigated here for the evolution of DCBH hosting haloes. The blue case on the left is the isolated, growing halo case and the pink shows an in-falling halo case. The star symbol represents the DCBH hosting proto-galaxy/galaxy while the dots represent a separate, central galaxy.}
   \label{fig:halodiagram}
\end{figure}

The formation of a seed BH through direct collapse is expected to take place in haloes within regions of high-intensity local LW radiation \citep[e.g][]{Agarwal2012b}. Recent studies have shown a local source of H$_2$-dissociating radiation from nearby quasars or PopII or PopIII stars is required \citep{Agarwal2016a} to provide the critical LW intensity. For this reason the properties of our model proto-galaxy are chosen to reflect those expected in proximity to a larger galaxy which formed at an earlier time. We assume the LW radiation field is sufficient to entirely dissociate molecular hydrogen in the proto-galaxy and the gas temperature is set to $T_{\rm g}= 8000$ K. 
Due to this assumed proximity there is a strong likelihood of a DCBH hosting halo to undergo a merger in its evolution \citep{Agarwal2014a}. To account for the variation in the growth of seed BH hosting dark matter haloes we will focus on two evolutionary paths reported in simulations \citep{Agarwal2014a} (see Figure~\ref{fig:halodiagram}): (1) an isolated, growing halo and (2) a halo that forms a BH before it becomes a satellite at some later in-fall redshift, $z_{\rm infall}$. The main difference between these paths is the rate at which new baryons enter the proto-galaxy.

As mentioned above, the cooling in DCBH hosting haloes must be limited to occur via atomic hydrogen \citep{Agarwal2012b}. This constrains the metallicity, virial temperature, and therefore the mass of the haloes. All modelled haloes therefore have $T_{\rm v}\gtrsim 10^{4}$ K and the gas is comprised of atomic hydrogen and helium. For simplicity we assume that the cooling timescale is very short and cooling occurs on a dynamical time of the halo, meaning gas accreted by the halo reaches the proto-galaxy on a halo dynamical time\citep{Dekel2009, Khochfar2011}. An upper limit on the mass of such a proto-galaxy is therefore the baryonic mass fraction of the halo.

\subsection{Halo Growth} \label{haloevo}

We here model the time evolution of the system composed of a gaseous disc, a stellar disc, a BH and a dark matter halo. The evolving model follows the growth of a galactic disc within an isolated growing host halo after the formation of a massive BH seed at some initial redshift, $z_{\rm i}$, down to redshifts where SMBHs have been observed $z_{\rm end} \sim 6.0$ \citep{Fan2006a, Mortlock2011}. The seed formation redshift for our fiducial model is $z_{\rm i}\sim 10$ \citep{Agarwal2012b} though a range of seed formation redshifts are possible \citep[e.g.][]{Begelman2006} which we will investigate below as well. We follow the evolution of isolated and in-falling DCBH hosting haloes (see Figure~\ref{fig:halodiagram}).

The total mass of the system, $M_{\mathrm{total}}$, is made up of the dark matter (DM) halo and the baryons that make up the massive BH and the galaxy disc. To fit with the conditions expected for the formation of a DCBH, the total initial mass, $M_{\mathrm{total, \;i}}$, is calculated by estimating the mass of an atomic hydrogen cooling halo at $z_{\rm i}$ \citep{Mo2010}. The baryon fraction is assumed to follow the universal value of $f_{\mathrm{b}} = 0.17$. Initially, the BH seed is given a mass in the range $M_{\mathrm{\bullet, \; i}} = 10^{4 - 6} \, \mathrm{M_{\odot}}$, and the remaining baryons make up the disc mass.

The total mass increases through cosmological accretion of mass onto the system. The following equation taken from \citet{Dekel2013} is used to calculate the growth of the system:
\begin{equation}
	M_{\rm total}(z) = M_{\mathrm{total, \;i}} \, \mathrm{e}^{-\alpha(z - z_{\mathrm{i}})}
	\label{eq:1}
\end{equation}
Two methods are used to calculate the halo growth parameter, $\alpha$. In general, $\alpha=3/2 \, s \, t_1 = 0.806$ where $s = 0.030 \, \mathrm{Gyr^{-1}}$\citep{Dekel2013} and $t_1 = 2/3 \, \Omega_{m}^{-1/2} \, H_{0}^{-1} \sim 17.9 \, \mathrm{Gyr}$. As an alternative growth rate, $\alpha = 0.586$ was found by fitting an exponential to the median of the model growth histories for the host halo of the CR7 DCBH from \citet[][see their Figure 4]{Agarwal2016a}.

The total mass is split into the dark matter halo $M_{\mathrm{200}}(z) = M_{\mathrm{total}}(z) \, (1 - f_{\mathrm{b}})$ and the baryons. The model assumes the disc and central BH comprise all the baryons in the system and that accreted baryons go directly onto the disc, conserving mass. The equilibrium solutions are used here rather than allowing the accreted material to reach the disc over a dynamical time as the latter required an extra step in the calculation while having little baring on the disc mass at later times and, therefore, the redshift at which the disc became unstable. Hence, the mass of the disc at any given time is the difference between the total baryonic mass and the mass of the BH.

When sufficient gas is available, the BH is assumed to grow at a constant Eddington fraction, $f_{\mathrm{Edd}}$, leading to the following equation for the growth of the BH:

\begin{equation}
	M_{\mathrm{\bullet}}(t) = M_{\mathrm{\bullet, \;i}} \, \exp\left(f_{\mathrm{Edd}} \frac{t - t_{\mathrm{i}}}{t_{\mathrm{Sal}}}\right)
	\label{eq:5}
\end{equation}

\noindent where $t_{\mathrm{Sal}} = 0.45 \, \eta/(1 - \eta)\; \mathrm{Gyr}$ is the Salpeter timescale and $\eta$ is the radiative efficiency which we assume throughout as $\eta=0.1$ \citep{King2008}. The BH is assumed to only accrete gas. With a high Eddington fraction the BH accretion rate can exceed the baryonic growth rate of the halo at late times. This results in a decrease in the gas mass of the disc. If the gas mass drops to zero, the BH accretion rate will be limited to the baryonic growth rate of the halo.

\subsection{Star Formation and Stellar and Gaseous Disc} \label{evodisc}

Initially, we assume a gaseous disc which is fed by the net accretion of gas resulting from the difference in the gas accreted onto the disc and the BH. The growth of the disc translates into a growth in the gas surface density such that the central surface density becomes a function of time.

\begin{equation}
	\Sigma_{\rm g,\,0}(t) = \frac{M_{\rm g}(t)}{2\pi R_{\rm d}(t)^2}
    \label{eq:sigg0}
\end{equation} 

After the onset of star formation in the proto-galaxy, some of the gas is converted into stars. The SFR is calculated using the method discussed in section~\ref{starform}. For simplicity, stars are assumed to remain on circular orbits where they form in the disc. The stars therefore follow a different surface density profile to the gas and the total disc surface density is simply the sum of the stellar and gas surface densities.

\begin{equation}
	\Sigma_{\mathrm{d}}(R,t) = \Sigma_{\mathrm{g}}(R,t) + \Sigma_{\mathrm{\star}}(R,t)
	\label{eq:7}
\end{equation}

For the purpose of our model we neglect feedback from stars and the accreting BH and note that the stellar mass is an upper limit on what could be expected. The velocity dispersion is assumed to be dominated by the sound speed of the gas ($c_{\rm s}\sim 10\, \mathrm{km/s}$ with $T_{\rm g}=8000$ K). However, including feedback effects should lead to an increase in the velocity dispersion of $\sim 10 \, \mathrm{km/s}$ \citep[see, e.g.][]{Wada2002, Dib2006, Agertz2009} due to supernovae after a few Myr \citep{Schaerer2002}, driving outflows and suppressing star formation.

\begin{table}
	\centering
	\caption{Table of model parameters.}
	\label{tab:para_table}
	\begin{tabular}[h]{l|l|l|l} 
		\hline
		Parameter & Definition & Fiducial (Range)\\
		\hline
		$M_{\rm \bullet, \,i}$ & BH seed mass ($\mathrm{M_{\odot}}$) & $10^{6}$ ($10^{4 - 6}$) \\
		$f_{\rm Edd}$ & Eddington fraction & 0.25 ($0.0 - 1.0$)\\
		$\alpha$ & Halo growth parameter & 0.806 ($0.806, 0.586$)\\
		$z_{\rm i}$ & Seed formation redshift & 10.0 ($20.0 - 10.0$)\\
		$z_{\rm infall}$ & In-fall redshift & 0.0 ($10.0, 8.5, 7.0, 0.0$)\\
	\end{tabular}
\end{table}

\subsection{Fiducial Case} \label{fiducialevo}

Table~\ref{tab:para_table} summarises the parameters discussed above with their fiducial values and the relevant ranges used. For our fiducial evolving model we make comparisons between cases both with and without a BH and with and without BH accretion. For the growth of the halo and the disc we assume an accretion rate in line with \citet{Dekel2013}. Using the lower accretion rate and a $T_{\rm v}\sim 10^{4}$ K halo at $z_{\rm i}=10$ would result in a system where the disc was never massive enough to be unstable prior to $z = 6$, independent of the BH mass\footnote{At higher formation redshifts, $z_{\rm i}\sim20$, a $T_{\rm v}\sim 10^{4}$ K halo can form a disc capable of becoming unstable at later times even for the assumed lower accretion rates.}. Our fiducial value for the seed mass is $M_{\bullet}=10^6 \, M_{\odot}$ and we assume an Eddington fraction of $f_{\rm Edd}=0.25$ for the accreting BH case.

\begin{figure}
	\includegraphics[width=\columnwidth]{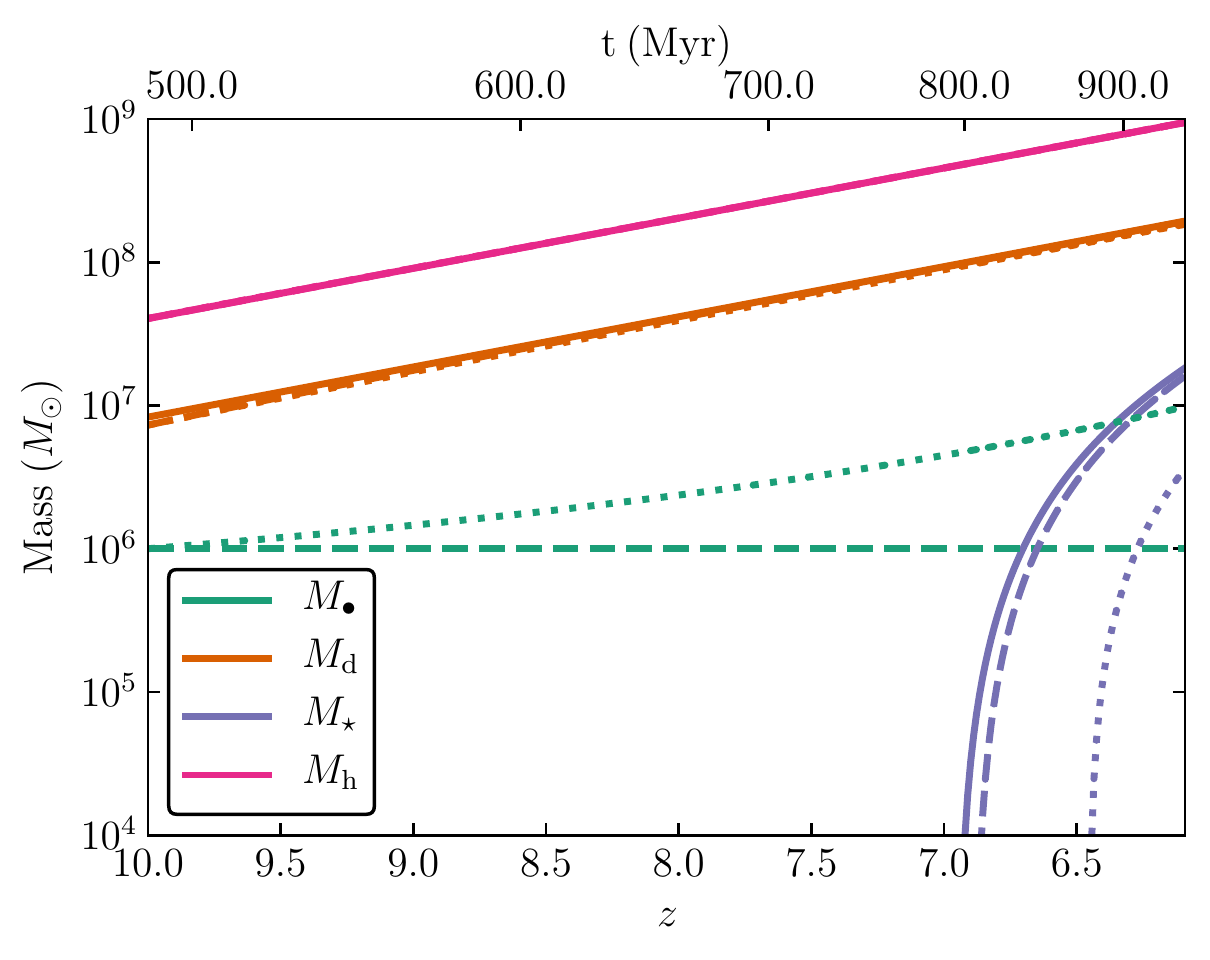}
    \caption{The mass evolution of each of the system components for the fiducial cases. The solid line is the case with no BH, the dashed line is for the case of a BH with a constant mass of $M_{\bullet} = 10^6 \, M_{\odot}$, and the dotted line is for the case with a BH growing from an initial mass of $M_{\bullet} = 10^6 \, M_{\odot}$ with $f_{\rm Edd} =0.25$.}
    \label{fig:fid_mevo}
\end{figure}

Figure~\ref{fig:fid_mevo} shows the evolution with redshift of the mass of each component of the model for our three fiducial cases (no BH, non-accreting BH, accreting BH). The stellar mass evolution varies between the different models. The case without the BH has the largest stellar mass at all redshifts after the onset of star formation while the accreting BH has the lowest. As the accreting case has the most massive BH it will have the longest star formation timescales and the highest stable disc fraction, leading to lower star formation rates and hence lower stellar masses. Furthermore, the higher BH mass leads to a delay in the onset of star formation. The higher BH changes the $Q_*$ profile such that $Q_{*,\,\rm min}$ is higher for a given disc mass and therefore the critical mass of the disc required for it to become unstable is higher. In the constant BH mass case, this higher critical mass requirement delays the time at which the disc is first unstable as each model has the same cosmological accretion rate. In fact this delay is further enhanced in the accreting BH case as the net growth rate of the disc will be reduced.

\subsection{Star Formation Rate Surface Density Profile}

\begin{figure}
	\includegraphics[width=\columnwidth]{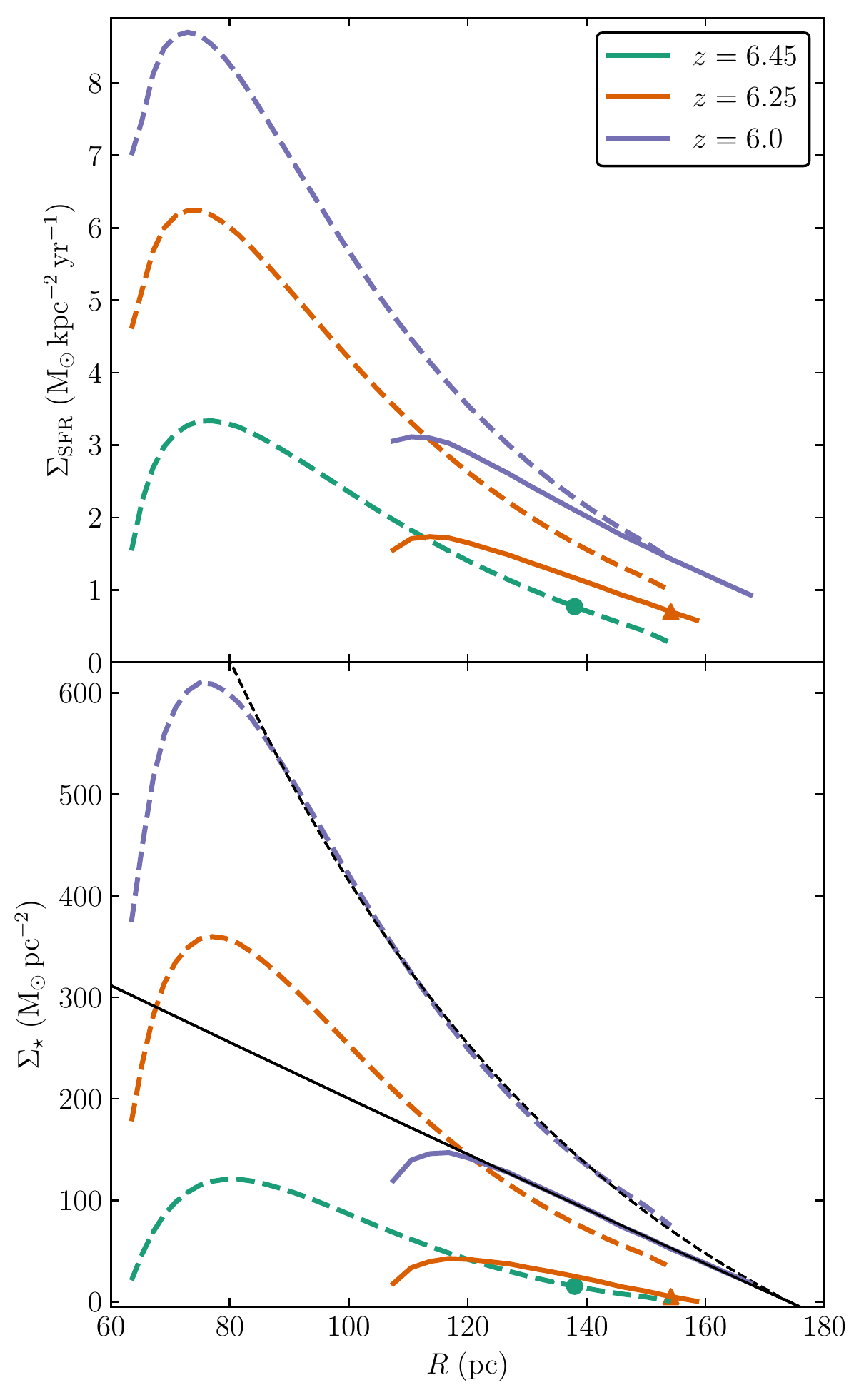}
	\caption{The radial profiles of the SFR surface density and the stellar surface density at three snapshots during the evolution of two models of disc galaxies. The solid lines represent the fiducial model with a growing BH and the dashed line is the case with no BH.  The first point in time ($z=6.45$) is taken immediately prior to the onset of star formation in the BH case and the final snapshot ($z=6.0$) is at the end of the calculation. In most cases, all star formation is taking place within the scale radius of the disc. Where this is not the case a marker of the corresponding colour indicates the disc scale length. The disc scale radius at each point in time is as follows: $R_{\rm d}=143.2$ pc for $z=6.45$, $R_{\rm d}=155.6$ pc for $z=6.25$, and $R_{\rm d}=167.6$ pc for $z=6$. Exponential surface density profiles, $\Sigma_{\star} \propto \exp\left (-a\, R/R_{\rm d} \right )$, were fitted to the stellar surface density profiles at $z=6$ (shown in black). At this redshift the fit parameter was found to be $a=0.51$ (with a turn over starting at $r=84$ pc) and $a=15.1$ (with a turn over at $r=124$ pc) in the BH and no BH cases respectively.}
    \label{fig:sigsfrprofile}
\end{figure}

The change in the Toomre and tidal parameter radial profiles due to the presence of a BH has an affect on the SFR in the disc. Figure~\ref{fig:sigsfrprofile} shows how the SFR surface density evolves in the model with and without a BH. The region where star formation takes place in the model is shifted outward in the cases with a BH compared to the one without. Over time the SFR increases throughout the unstable region. This is expected in our model as the formation of stars in a region increases the stellar surface density while the corresponding decrease in gas density is spread out throughout the disc. Meanwhile more gas is accreted through cosmological accretion and $R_{\rm d}$ increases as the halo grows. This means even as the gas density profile is stretch out there is an overall increase in the total surface density in a region undergoing star formation and this increases the SFR in that region (see Equation~\ref{eq:sf9}). The lack of feedback effects or any momentum and mass transfer in the stellar disc results in runaway star formation. As the BH is allowed to accrete gas there is further increase in the difference in the SFR at late times as the gas surface density is reduced. The differences in the stability profiles and gas density means the SFR and stellar mass surface densities are higher at each point in time and at each radius in the case without the BH. Looking at the lowest redshift, the presence of the BH has resulted in a decrease in the width of the annulus of the disc where stars can form by $\sim 1/3$. The resulting ring of stars occupies this same smaller region and the inner region is void of stars, effectively creating a hole in the galaxy stellar disc by enlarging the central region void of stars from $\sim64$ pc to $\sim110$ pc. The change in the $Q_*$ profiles does result in a fractional increase ($9\%$) in the outer radius of the star formation region but this only has a minor effect on the total SFR of the system. 

\subsection{Evolution of the Star Formation Rate}

\begin{figure}
	\includegraphics[width=\columnwidth]{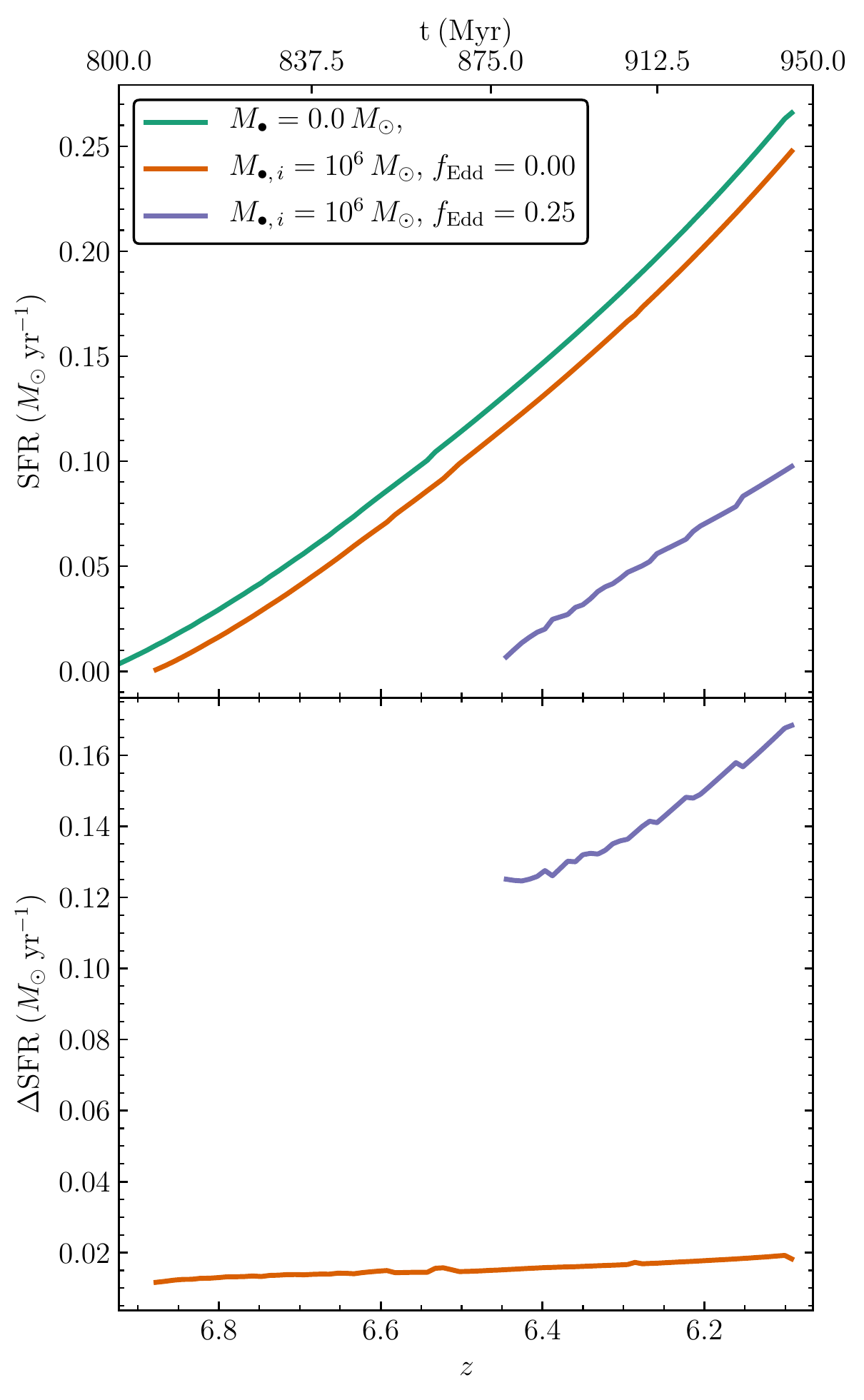}
    \caption{The evolution of the star formation rate for the fiducial cases and the difference between the cases with and without a BH. $\Delta \mathrm{SFR} = \mathrm{SFR}_{\rm no\,BH} - \mathrm{SFR}$.}
    \label{fig:fid_sfr}
\end{figure}

The radially integrated SFR in Figure~\ref{fig:fid_sfr} shows the difference in the total SFR over time. The SFR of the no BH case is highest at all redshifts after the onset of star formation and the accreting BH case results in the lowest. As we go forward in time we see that the difference between the SFRs increases. As the stellar density increases in the unstable region of the disc, the gas mass will continue to be spread across the total disc profile and the total density will increase within the unstable region, resulting in a local increase in the SFR (see Equation~\ref{eq:sf9}). As the model does not follow stellar migration, the stellar mass is not redistributed and the SFR simply increases the surface density and so on, leading to a run away effect until the gas density reaches the star formation threshold value. This means, within the model, once a galaxy has a higher SFR and stellar mass it becomes hard for another model to catch up, unless the gas is used up less efficiently due to the presence of a BH.

Note the inclusion of feedback from accretion onto the BH would have the potential to regulate the star formation in the disc further. The resulting heating and ejection of gas could further stabilise the disc against star formation but also regulate the growth of the BH \citep{Latif2018}. If the BH were to maintain the assumed growth rate, the star formation rate would be expected to decrease due to the decrease in the gas surface density and the increase in the gas temperature. The onset of star formation would be delayed and once the disc did reach instability it would have a lower unstable fraction, lowering the total SFR.

\begin{figure}
	\includegraphics[width=\columnwidth]{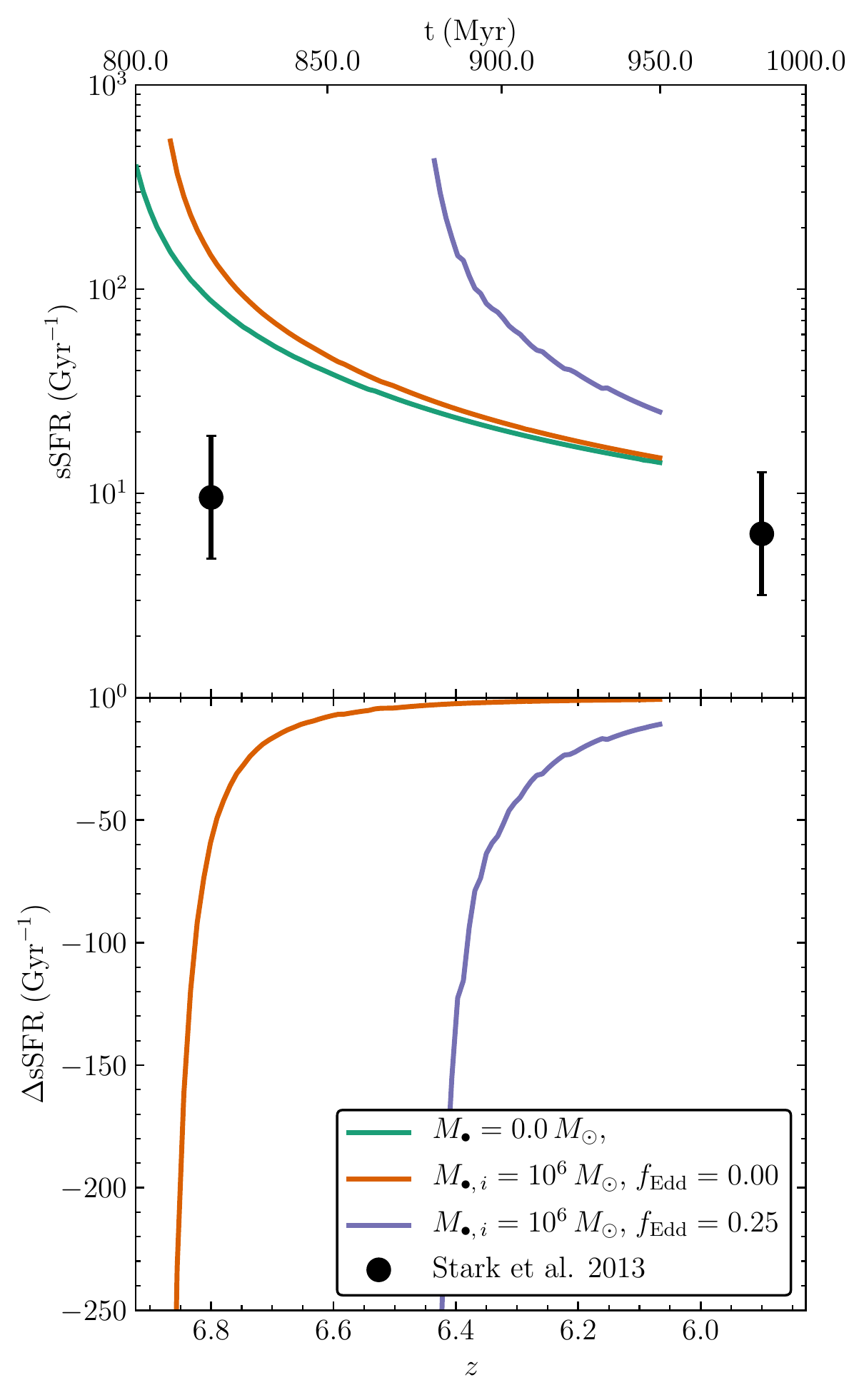}
    \caption{The evolution of the specific star formation rate for the fiducial cases and the difference between the cases with and without a BH. $\Delta \mathrm{sSFR} = \mathrm{sSFR}_{\rm no\,BH} - \mathrm{sSFR}$.}
    \label{fig:fid_ssfr}
\end{figure}

The evolution of the specific star formation (sSFR) rate of each of the models (Figure~\ref{fig:fid_ssfr}) shows interestingly that the sSFR is higher for the higher BH mass cases. The rise in $Q_{\rm Toomre}$ and $Q_{\rm tidal}$ due to the BH decreases the SFR and therefore a significant decrease in the stellar mass over time, resulting in an increase in the sSFR. Indeed, without the BH the sSFR is lower at early times as the stellar surface density will be significantly larger due to the difference in the time at which star formation can first occur in the disc. As the system progresses the stellar masses become more comparable and the difference in the sSFR decreases.

When compared to observations \citep{Stark2013}, we find our model sSFR is greater by a factor of $~10$ at $z=6.8$, though the lower mass BH and no BH cases appear to be following a trend which would agree with the $z=5.9$ data point. However, the relationship between SFR and stellar mass has a large scatter and the slope varies with stellar mass \citep{Whitaker2014}, meaning large deviations from this median value in sSFR for individual galaxies is to be expected, particularly at low masses. Indeed, our findings suggest DCBH hosting galaxies should generally have a higher sSFR, providing a possible tool for identifying candidate DCBH hosts.

\subsection{Onset of Star Formation}
\begin{figure*}
\begin{center}$
\begin{array}{cc}
\includegraphics[width=\columnwidth]{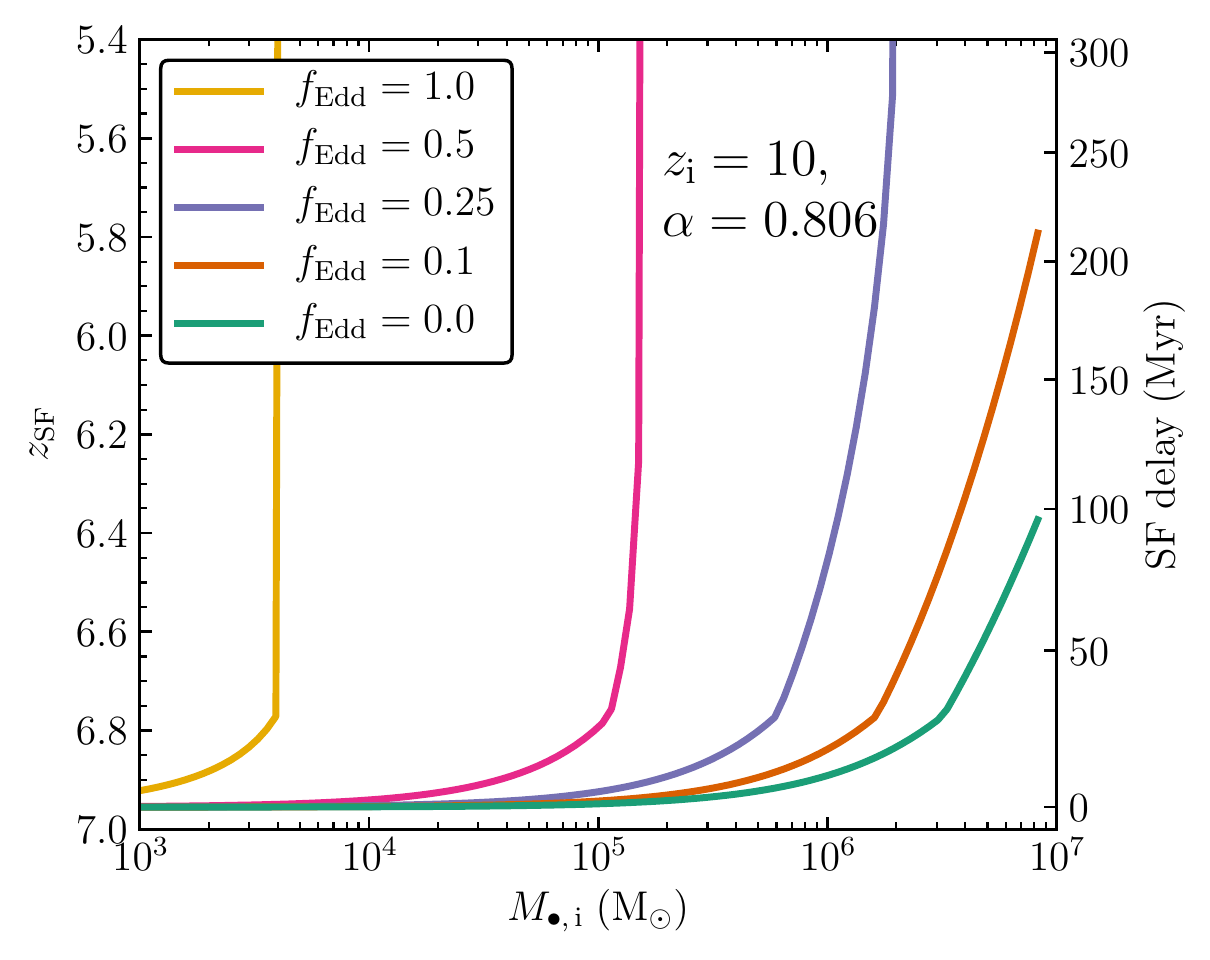} &
\includegraphics[width=\columnwidth]{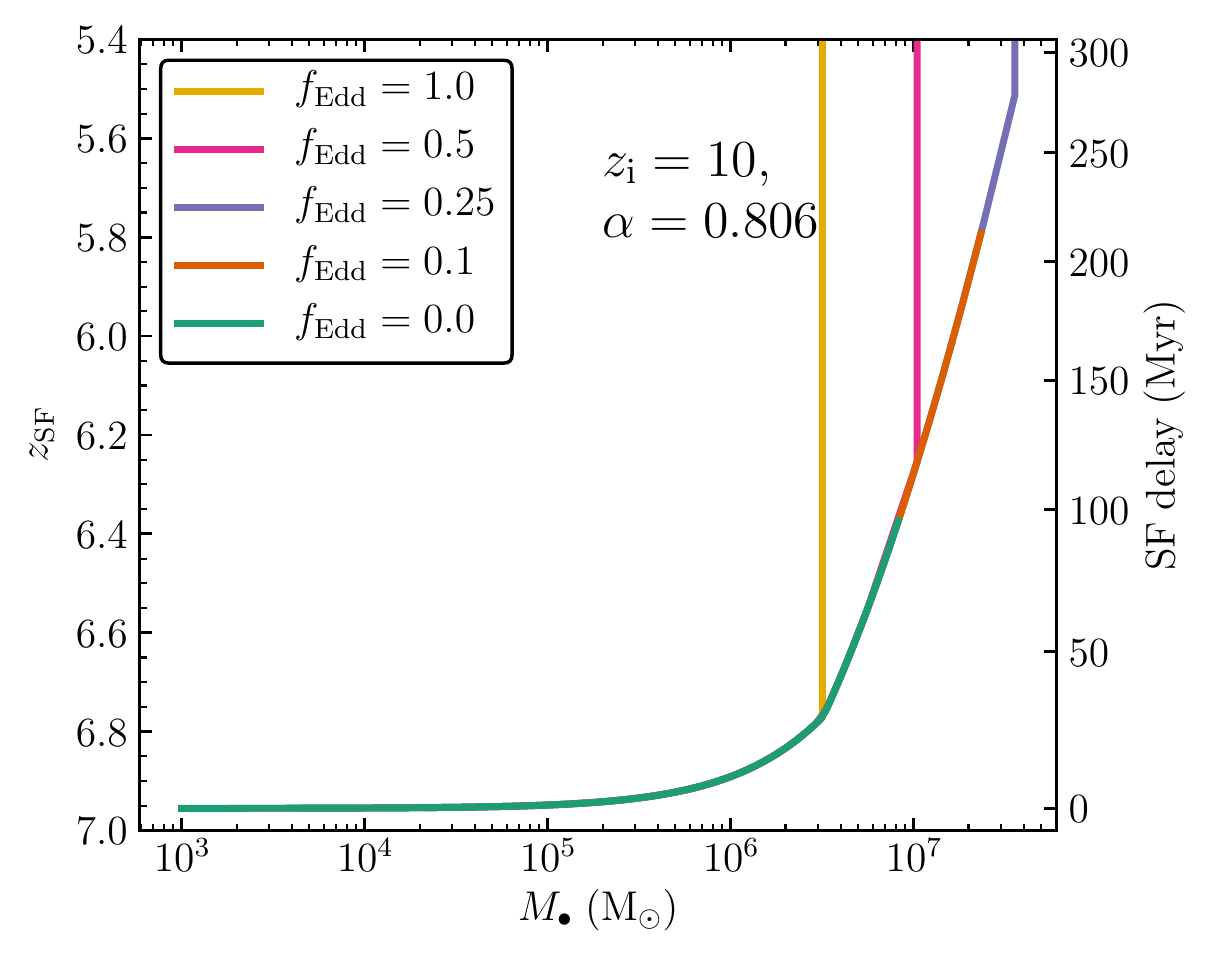}\\
\includegraphics[width=\columnwidth]{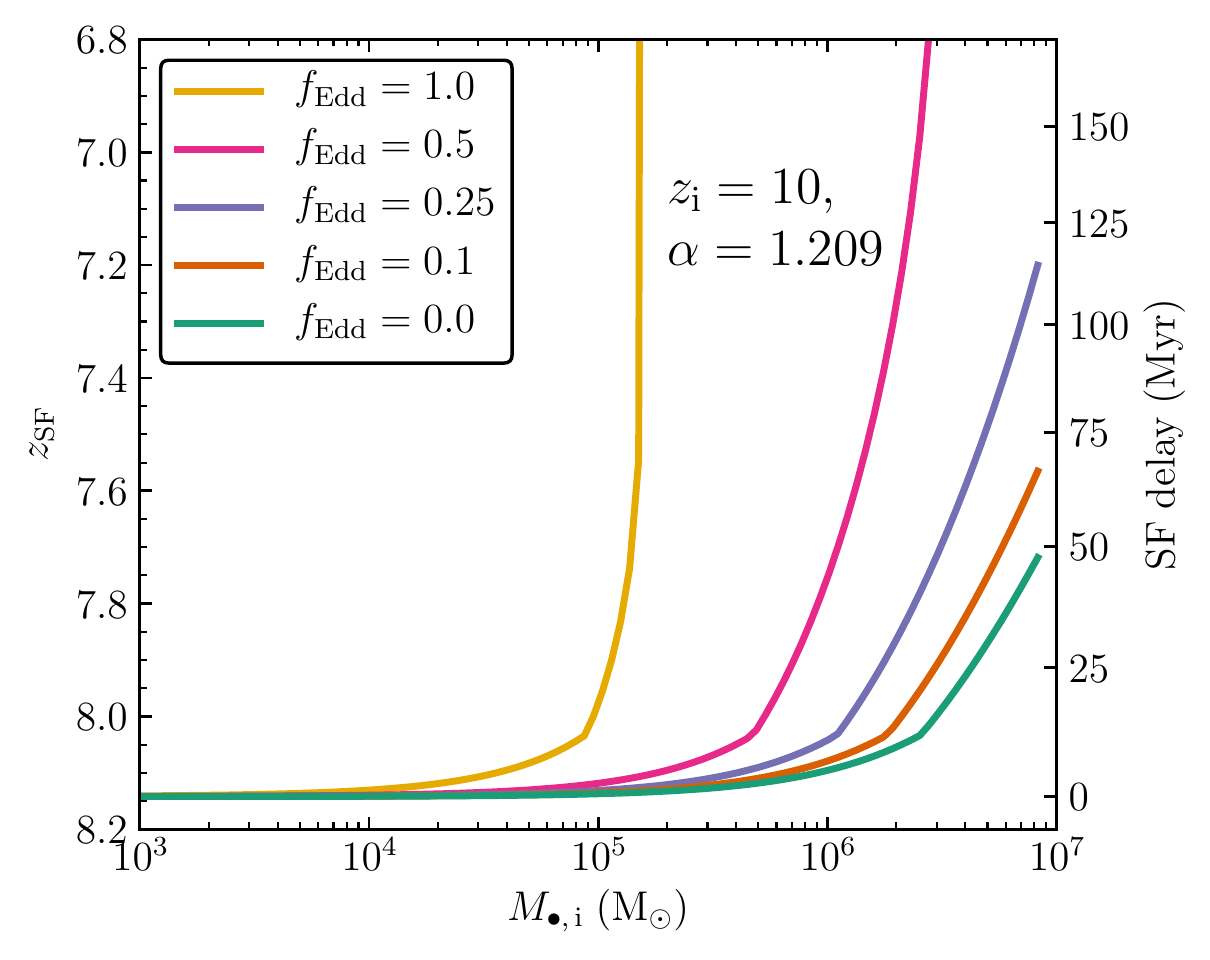} &
\includegraphics[width=\columnwidth]{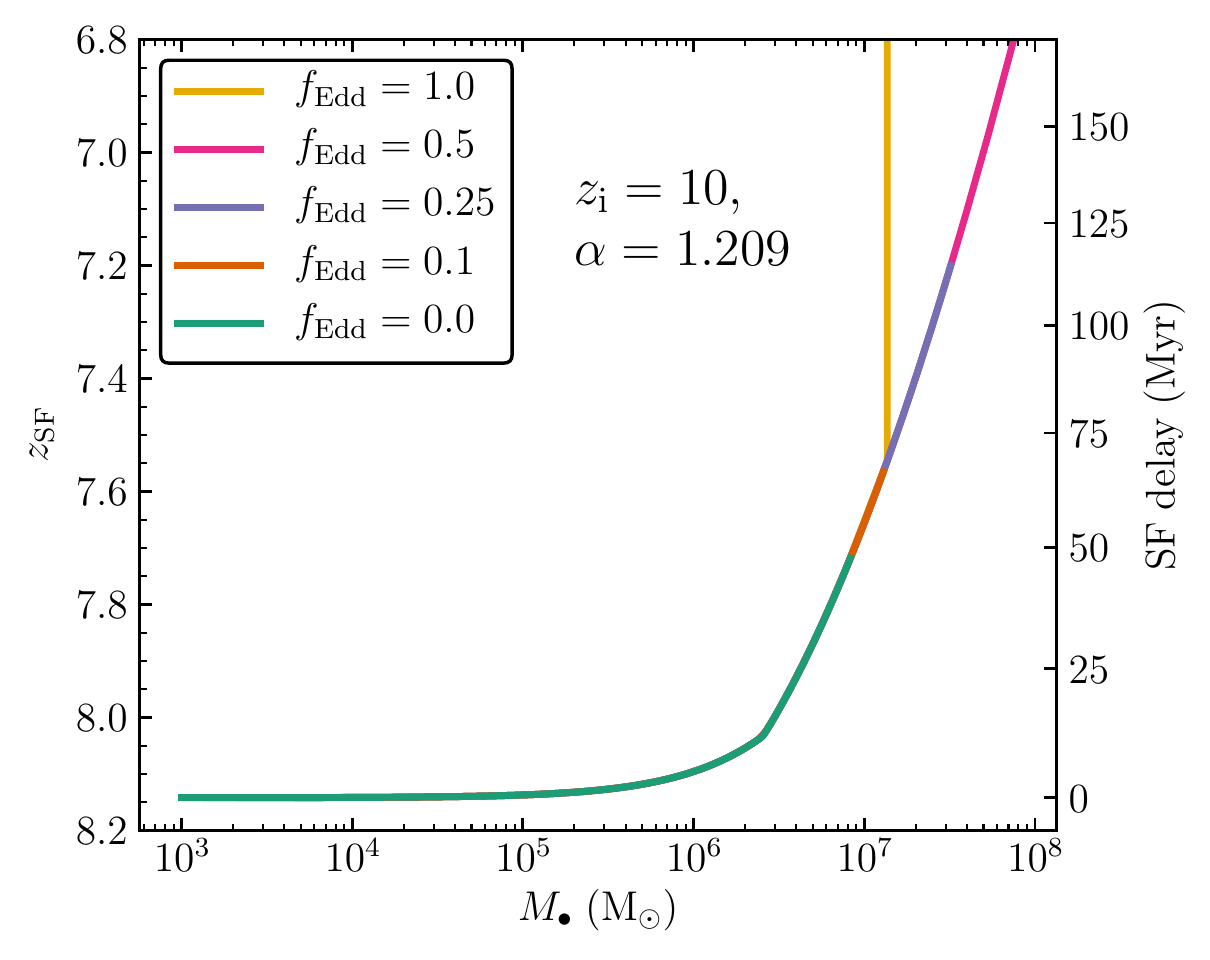}\\
\includegraphics[width=\columnwidth]{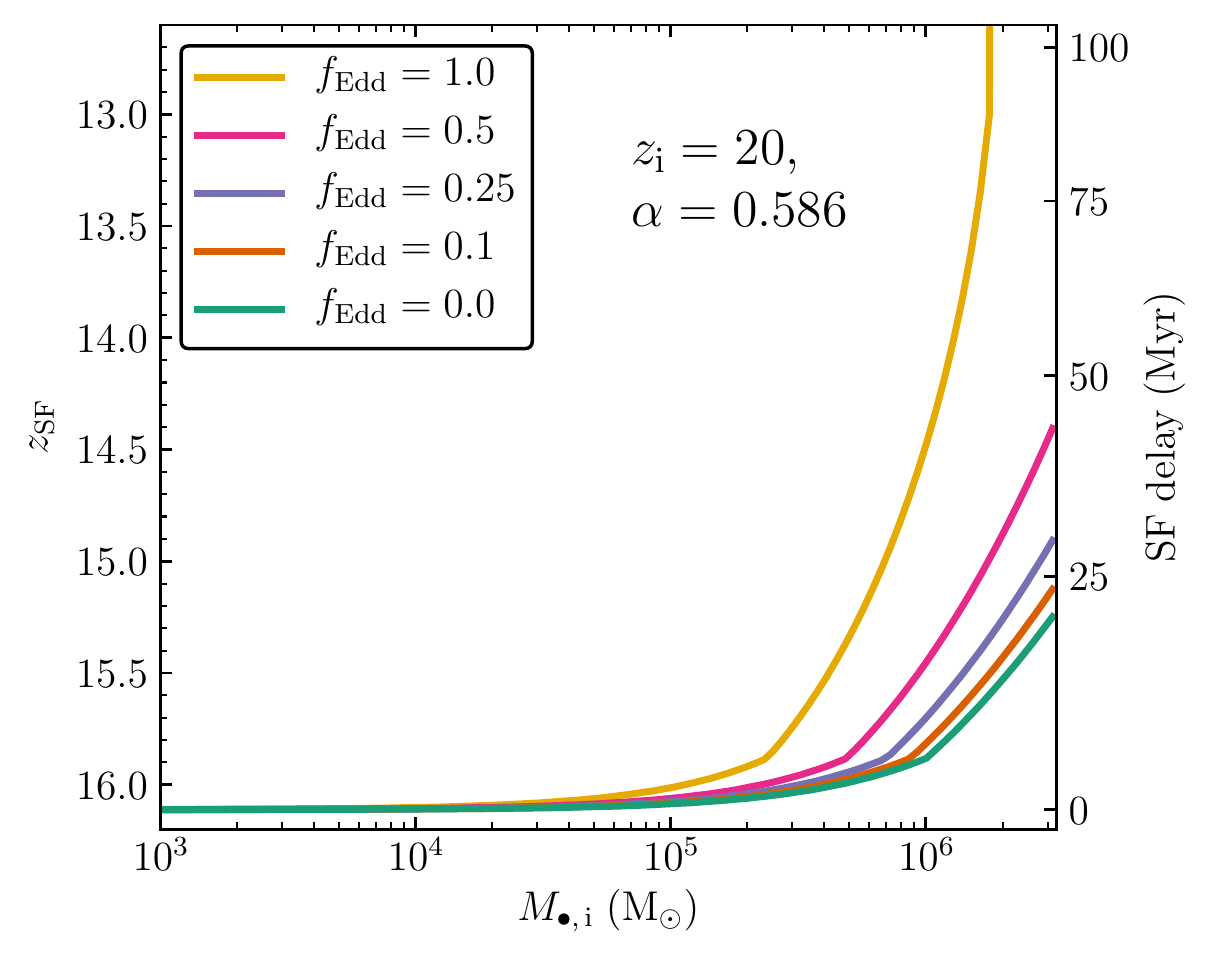} &
\includegraphics[width=\columnwidth]{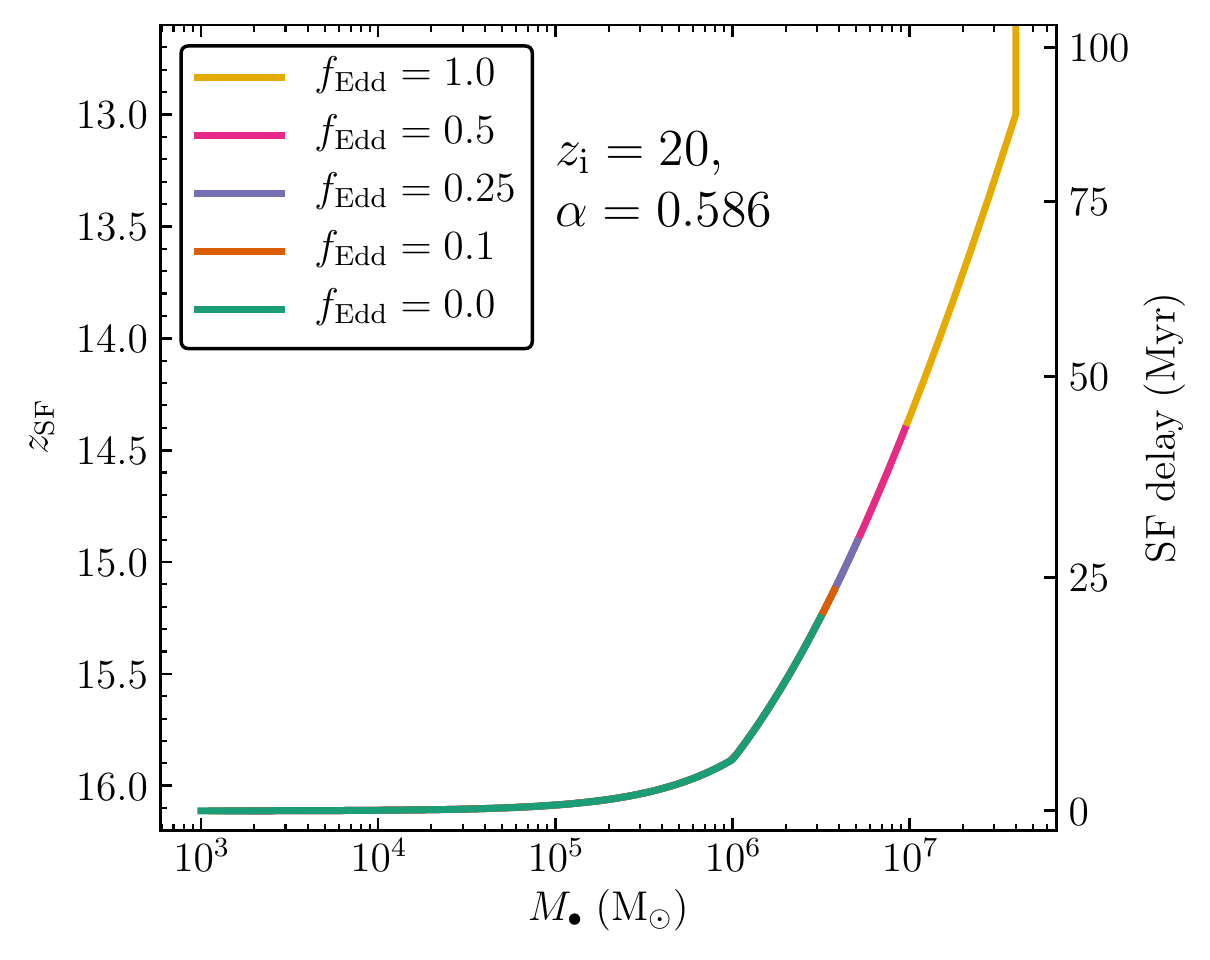}
\end{array}$
\end{center}
\caption{The left panels show the redshift at which the modelled disc first becomes unstable (i.e. $Q_{*\,\rm min}=1$) and is able to form stars ($z_{\rm SF}$) as a function of the initial BH seed mass ($M_{\bullet, \, i}$) for different fractions of the Eddington limit accretion rate ($f_{\rm Edd}$). The right panels show the same except with the BH mass at $z_{\rm SF}$ on the x-axis. The time difference between the onset of star formation with and without a BH is also shown on the right hand side to indicate the delay caused by the stabilising effect of the BH. The upper limit of the initial BH mass range is the total baryonic mass of the atomic hydrogen cooling halo at the DCBH formation redshift of $z\sim10$ for the top two rows of panels and $z\sim20$ for the bottom. The growth rate of the halo follows Equation~\ref{eq:1} with $\alpha=0.806$ for the top panels, $\alpha=1.209$ for the middle and $\alpha=0.586$ for the bottom. The $f_{\rm Edd} = 0.25 - 1$ lines in the top panels each reach a maximum seed mass above which the disc will never become unstable and be able to form stars. In the bottom panels this is only seen for the $f_{\rm Edd} = 1$ line.}
\label{fig:zsf}
\end{figure*}

The top panels of Figure~\ref{fig:zsf} show how the redshift at which star formation first occurs in the model depends on the seed mass and the growth rate of the BH for an atomic hydrogen cooling halo that forms a DCBH at $z=10$ while the remainder of its baryonic mass goes into making a disc. For the case of Eddington limited accretion ($f_{\rm Edd} = 1$) even the lowest mass in the estimated range of the DCBH masses, $M_{\bullet,i}=10^4 \; M_{\odot}$, results in a disc that will never undergo star formation. Yet, high accretion rates close to the Eddington limit are required for even the most massive DCBHs at $z\sim10$ to reach the $M_{\bullet}\sim10^9\, \mathrm{M}_{\odot}$ by $z\gtrsim6-7$ as observed \citep{Fan2006a,Mortlock2011}. This indicates DCBH formed and grew into SMBHs in separate progenitors from their eventual host galaxies, in order for these massive quasars to be observed within massive galaxies at $z\sim6$. In fact, for the upper limit of the DCBH mass range, $M_{\bullet,i}=10^6 \; M_{\odot}$, star formation is inhibited for the $f_{\rm Edd} = 0.5$ case and is delayed by $\sim 100$ Myr with $f_{\rm Edd} = 0.25$. Most notably at this formation redshift ($z_{\rm i}\sim10$), any combination of seed mass and growth rate which leads to the growth of a $M_{\bullet}\sim 10^9 \, \mathrm{M}_{\odot}$ SMBH by $z\sim6$ inhibits star formation in the host.

In the cases of no BH growth and $f_{\rm Edd} = 0.1$, the disc will eventually undergo star formation, even when the BH seed mass is at its maximum i.e. at the seed's formation it takes up the entire baryonic mass of the halo ($M_{\bullet,i}= M_{\rm b} = 7.82 \times 10^6 \; M_{\odot}$) and the disc mass is initially zero. The onset of star formation is delayed somewhat in these $M_{\bullet,i}= M_{\rm b}$ seed cases, with $f_{\rm Edd} = 0.1$ leading to a delay by $\sim 200$ Myr which is significant as this is around a fifth of the age of the universe at this epoch. Note in all models the BH mass never exceeds the total baryonic mass in the redshift range we investigate in the models where $z_{\rm SF}$ is defined.

The growth rate of the halo and therefore the disc greatly influences this result. The lower the growth rate of the halo the more delayed star formation will be. As highlighted above, the growth rate modelled for the DCBH hosting halo of CR7 by \citet{Agarwal2016a} is sufficiently low that with $z_{\rm i} = 10$ the surface density of the disc is never high enough for stars to form over the redshift range we investigate. However, at earlier formation times the role of the BH decreases as the growth rate of the halo becomes higher at larger redshift. With a formation redshift of $z_{\rm i}=20$, a BH with an initial seed mass of $M_{\bullet, \rm i}=10^6 \, \mathrm{M}_{\odot}$ growing at the Eddington limit will be unable to prevent star formation, only delaying the onset by $\sim40$ Myr (see bottom panels of Figure~\ref{fig:zsf}).

Similarly to increasing the formation redshift, increasing the growth rate of the halo decreases the influence of the BH. \citet{Genel2008} find a scatter in the growth rate of DM haloes which they approximate as $\approx \left \langle \dot{M}_{\rm DM} \right \rangle \left (2.5/(1+z) \right )^{0.2} $ where $\left \langle \dot{M}_{\rm DM} \right \rangle$ is the mean halo growth rate. Assuming the growth rate is a linear function of the halo mass, in line with Equation~\ref{eq:1}, leads to $\sigma_\alpha \approx \left \langle \alpha \right \rangle \left (2.5/(1+z) \right )^{0.2}$. The middle panels of Figure~\ref{fig:zsf} shows the case for a formation redshift of $z_{\rm i} = 10$ with a growth rate of $\alpha = 1.209$, $\times 1.5$ the fiducial rate and within the 1$-\sigma_\alpha$ scatter at $z=6$. With this case the disc rapidly becomes more massive than the fiducial case and therefore becomes unstable much earlier. The BH mass required to keep the disc stable increases; it has to grow much faster to keep up with the disc and prevent star formation. This is illustrated by looking at a $M_{\bullet, \rm i}\sim 2\times 10^5 \, \mathrm{M}_{\odot}$ seed case. Growing at the Eddington limit, such a seed does prevent star formation, however, this is achievable at the same growth rate by a seed with a mass 20 times smaller at $M_{\bullet, \rm i}\sim 5\times 10^3 \, \mathrm{M}_{\odot}$ in the fiducial case. Furthermore, a $M_{\bullet, \rm i}\sim 2\times 10^5 \, \mathrm{M}_{\odot}$ seed in the fiducial case is capable of preventing the onset of star formation growing at half the BH accretion rate.

\begin{figure*}
\begin{center}$
\begin{array}{cc}
\includegraphics[width=\columnwidth]{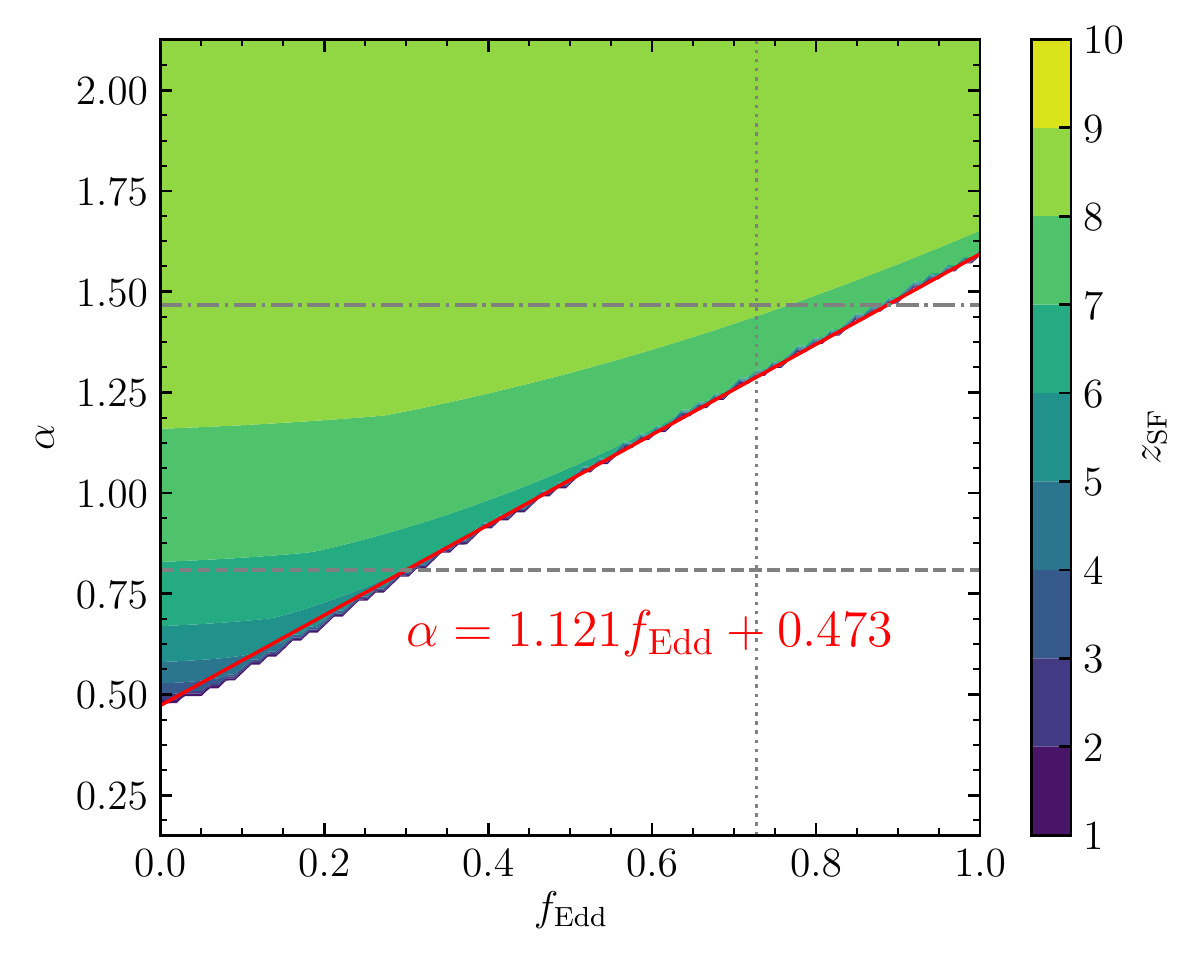} &
\includegraphics[width=\columnwidth]{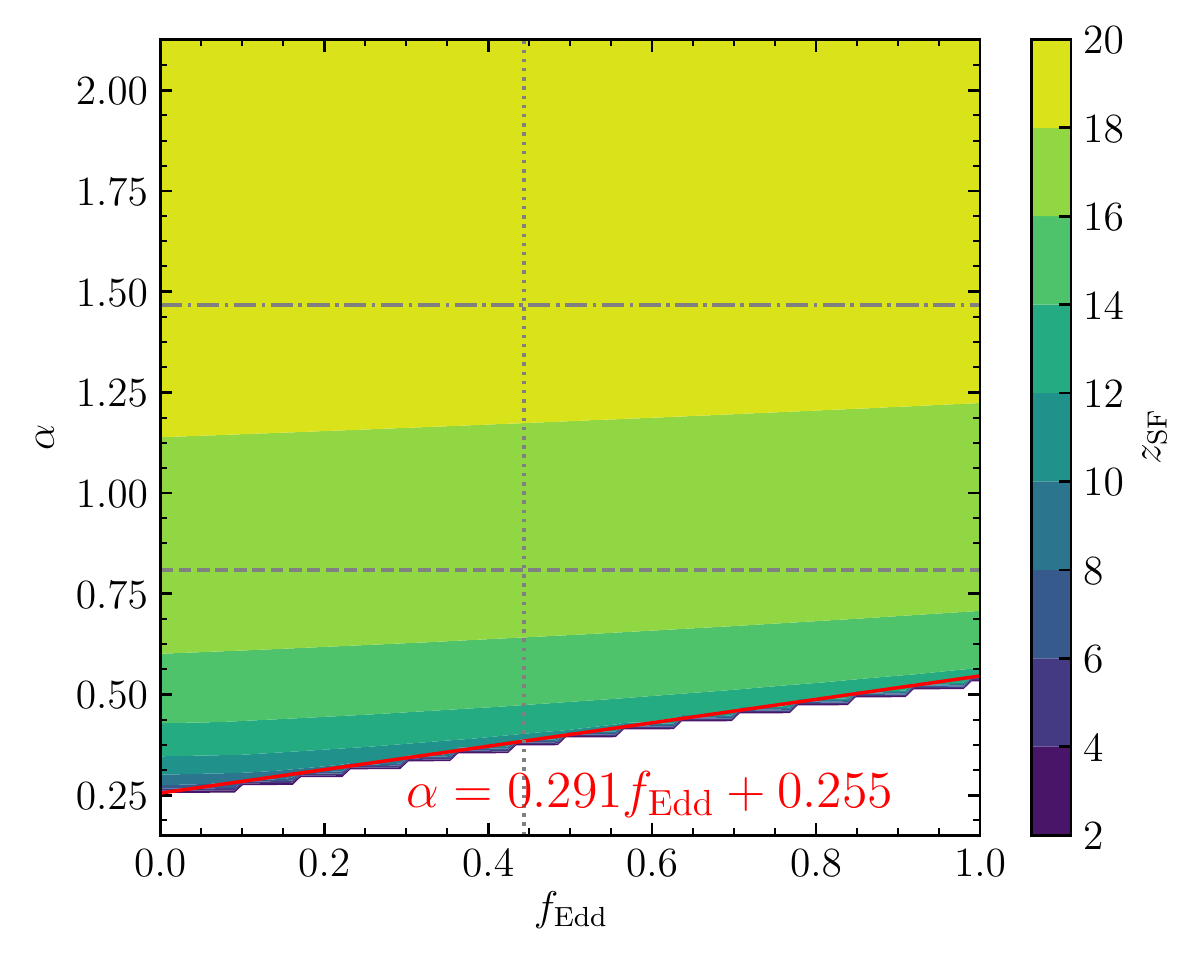}
\end{array}$
\end{center}
\caption{The variation of $z_{\rm SF}$, the redshift at which star formation can first occur, with $\alpha$ and $f_{\rm Edd}$, the accretion rate parameter of the halo and the BH Eddington fraction respectively. Where no value for $z_{\rm SF}$ is shown, the system will never undergo star formation. The initial seed mass is $M_{\bullet, \rm i}=10^6 \, \mathrm{M}_{\odot}$. The formation redshift is $z_{\rm i}=10$ in the left panel and $z_{\rm i}=20$ in the right panel. The range in $\alpha$ shown is from the lower 1$-\sigma_{\alpha}$ limit to the upper 2$-\sigma_{\alpha}$ limit at $z=6$. The average halo accretion rate parameter, and our fiducial value, $\alpha=0.809$ is shown as the grey, dashed line. The upper 1$-\sigma_{\alpha}$ limit of $\alpha=1.467$ is shown as the grey dot-dash line. The grey, dotted line represents the minimum Eddington fraction for which a $M_{\bullet, \rm i}=10^6 \, \mathrm{M}_{\odot}$ seed BH will reach $M_{\bullet}=10^9 \, \mathrm{M}_{\odot}$ by $z=6$. A best-fitting line for $f_{\rm Edd}>0.01$ is shown in both cases for the critical values of $\alpha$ and $f_{\rm Edd}$ where the model transfers from becoming unstable at some redshift to never being able to form stars.}
\label{fig:alpha_fedd_zsf}
\end{figure*}

The interplay of the halo and BH growth rates is well summarized in Figure~\ref{fig:alpha_fedd_zsf}. The figure shows how the onset of star formation varies with the growth rate of the halo and the BH and also depends on the formation redshift. $z_{\rm SF}$ is calculated as a function of the halo growth parameter, $\alpha$, and Eddington fraction for the same seed mass of $M_{\bullet, \rm i}=10^6\, \mathrm{M}_{\odot}$ at formation redshift $z_{\rm i}=10$ and $z_{\rm i}=20$. The range in $\alpha$ shown is from the lower 1$-\sigma_{\alpha}$ limit to the upper 2$-\sigma_{\alpha}$ limit at $z=6$, where $\sigma_{\alpha}$ was calculated using the approximation from \citet{Genel2008} as outlined above. In the $z_{\rm i}=10$ case, a significant fraction of the parameter space results in a model that is unable to ever form stars, particularly at higher Eddington fractions. Above the Eddington fraction at which the BH reaches $M_{\bullet}=10^9 \, \mathrm{M}_{\odot}$ at $z=6$ ($f_{\rm Edd}=0.728$), a higher than average halo growth parameter is required for star formation to occur. At even higher BH accretion rates, $f_{\rm Edd}\gtrsim0.9$, only haloes growing more than 1$-\sigma_{\alpha}$ faster than the average growth rate are sufficient. However, with $z_{\rm i}=20$, only the models with a slower than average halo growth rate, $\alpha \lesssim 0.8$, have a significantly delayed onset of star formation.

\begin{figure}
	\includegraphics[width=\columnwidth]{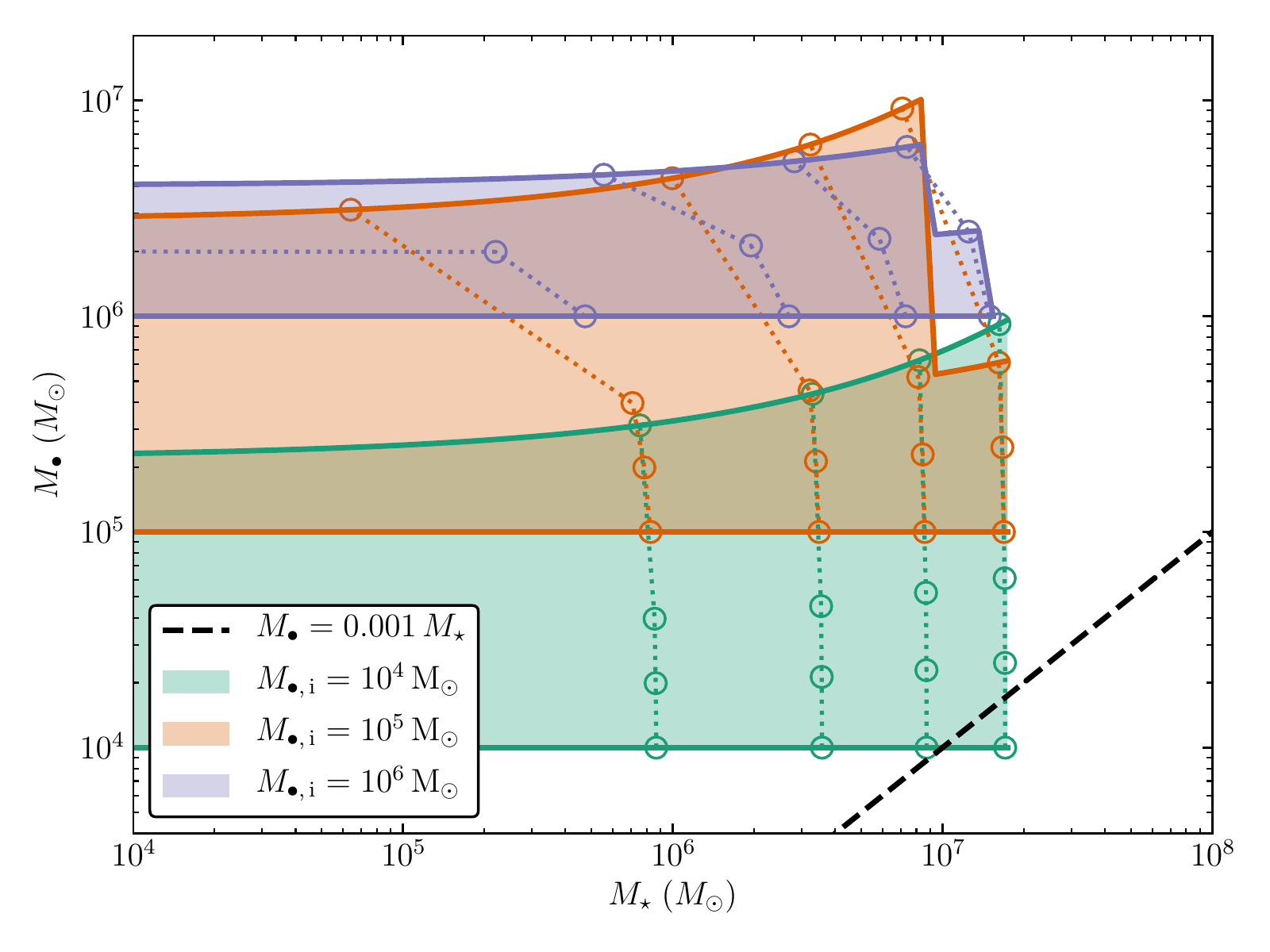}
	\caption{The evolution of the total stellar mass and BH mass was calculated for a range of growth rates ($f_{\rm Edd} = 0.0,\,0.1,\,0.2,\,0.5,\,1.0$; note $f_{\rm Edd}=1.0$ cases are not shown) and seed masses ($M_{\bullet,\rm \,i}=10^6 \, \mathrm{M_{\odot}},\,10^5 \, \mathrm{M_{\odot}},\,10^4 \, \mathrm{M_{\odot}}$). The coloured regions represent the spread in the evolution for different accretion rates given the same initial seed mass. Each model starts from the left hand side of the plot (with a stellar mass of zero) and once the onset of star formation is reached the evolution traces from left to right. The BH accretion rates shown here are limited to $f_{\rm Edd} \le 0.2$, $f_{\rm Edd} \le 0.5$, and $f_{\rm Edd} \le 0.5$ in the $M_{\bullet,\rm \,i}=10^6 \, \mathrm{M_{\odot}}$, $M_{\bullet,\rm \,i}=10^5 \, \mathrm{M_{\odot}}$, and $M_{\bullet,\rm \,i}=10^4 \, \mathrm{M_{\odot}}$ seed mass cases respectively as the higher accretion rates entirely prevent star formation. The dashed line represents $M_{\bullet}= 10^{-3}\, M_{\star}$ \citep[see, e.g.][]{Kormendy2013} and the coloured dotted lines connect points at the same redshift for each seed mass case. From left to right the lines correspond to $z=6.7,\,6.5,\,6.3,\,6.1$.}
    \label{fig:MbhMsevo}
\end{figure}

How strong an effect the BH has on the galaxy will depend on the growth rate of both the BH and the disc. Figure~\ref{fig:MbhMsevo} shows the evolution of the BH and stellar mass for different seed masses and accretion rates. The BH mass is initially significantly more massive but as the SFR is significantly larger than the BH accretion rate in these cases, the stellar mass quickly catches up with the BH mass. However, by the end of the calculation at $z=6$, only the lowest mass seed region reaches to the $M_{\bullet}\sim 10^{-3}\, M_{\star}$ line seen empirically at lower redshifts \citep{Haring2004, Kormendy2013, Reines2015}. This implies a BH cannot grow to lie on the BH-stellar mass relation at this point in cosmic time if the halo grows with an average growth rate. Indeed, earlier studies have indicated that systems reach the BH-stellar mass relation through periods of growth triggered by galaxy encounters \citep{Lamastra2010, Valiante2014}. The results of our model indicate a boost to the stellar mass is required possibly through mergers with evolved galaxies hosting only small or no BHs once the SMBH has grown. This suggests it is likely that these seeds are generated in satellites prior to falling into their host galaxies to lie on the relation. 

It is also thought that the empirical scaling relation of BH and stellar mass is linked to the interaction of AGN and star formation; the feedback attributed to AGN helps regulate the SFR and vice-versa \citep[see, e.g.][]{Silk1998, Gabor2010, Silk2013}, producing this correlation. Here we do not model the feedback from either the BH or stars yet it is not clear that the inclusion of feedback would resolve the discrepancy of the results with the empirical relation. This model predicts BH and galaxy masses that would place the model galaxies above the relation, meaning the BHs are too massive relative to their hosts. Yet the inclusion of feedback from stars to regulate the growth of BHs would be insufficient as our BHs gain most of their mass prior to the onset of star formation. The inclusion of feedback from accretion onto the BH itself would be expected to regulate both the growth of the BH and the SFR. If, however, the BH were able to grow at the assumed rate in the model, the heating of the disc due the BH growth would decrease the fraction of the gas able to form stars. The resulting decrease in the SFR would lead to such systems existing further above the BH-stellar mass relation.

\subsection{In-falling Host Halo}

The in-fall of the seed-BH-hosting halo to become a satellite of a more massive, central galaxy is modelled by cutting the growth of the halo. This reflects the starving of satellites as they are unable to accrete fresh material and existing material will either be used up or stripped. At a given redshift, $z_{\rm infall}$, the halo growth rate is set to zero and in turn accretion of fresh baryons stops. The disc mass only changes as the BH continues to accrete the remaining gas.

If the in-fall event happens prior to the onset of star formation the disc will never become unstable, due to the halt in the growth of the disc mass. This would result in a massive BH surrounded by a primordial gaseous disc (assuming the BH is unable to accrete all the gas), orbiting a central galaxy. For example, this would be true in our fiducial case if the in-fall occurs at a higher redshift than the onset of star formation at around $z\lesssim 7$. As the delay in onset of star formation increases with the mass of the seed BH, an in-fall event is more likely to occur prior to the onset of star formation in systems with more massive BHs.

In general, the SFR is maintained by the influx of gas to the disc. Cutting off this supply by having an in-fall after the onset of star formation leads to a rapid decrease in the gas density as the gas is converted into stars and accreted by the BH. The rate at which the SFR then decreases to zero will depend on the stellar mass and the BH accretion rate. Eventually, the stellar and BH masses will each reach a maximum and stop growing, starved by the lack of gas. For our fiducial model values for BH formation redshift and halo growth rate, only the models with lowest mass and slowest growing BH seeds can reach the $M_\bullet/M_\star\sim0.001$ relation by $z\sim6$ (Figure~\ref{fig:MbhMsevo}). Therefore, if the in-fall of the model galaxy were to take place at $z>6$, the resulting satellite would likely have an oversized BH relative to its stellar mass. This low mass satellite dominated by a massive BH could survive to lower redshifts due to the absence of further accretion of gas. This does however depend on the timing of the subsequent merger of the satellite with the central galaxy and, hence, the separation and relative masses of the merging galaxies. Indeed, the merging of massive BHs from the accretion of BH-dominated satellite galaxies could help form the SMBHs found in massive central galaxies at lower redshift \citep{Volonteri2003}, however, the time-scale for massive BHs to merge could be large \citep{Tremmel2015}.

A study by \citet{Agarwal2014a} identified haloes where DCBHs formed within a cosmological, hydrodynamical simulation. We highlight two of the cases they identified with a DCBH formation redshifts of $z_{\rm i}\sim 10$. One where the seed forms in a site close to one dominant galaxy and another where the formation in a clustered environment. In first case the DCBH host falls in to its largest neighbour at $z\sim8.5$, which is $\lesssim 200$ Myr after the formation of the seed. Within our model this would likely occur prior to the onset of star formation in the seed hosting halo. In the second case the seed hosting halo undergoes an in-fall at a later time, around $z\sim6$, well after the likely onset of star formation from our model. However, the environment of the cluster may play a stronger role in this scenario.

\section{Discussion \& Conclusions} \label{conclusions}

We use an analytical model to investigate the effect of a DCBH seed on the stability of proto-galaxy discs and the resulting suppression of star formation. We look at how the Toomre and tidal stability parameters profiles of an exponential disc change due to the presence of a BH in the centre of the system and link the stability of the disc to the star formation rate. We show how the BH has a gravitationally stabilising effect on the inner region of the disc which increases the star formation timescale locally and limits the region of the disc where star formation can occur, decreasing the modelled SFR. We also model the growth of a galaxy around a seed BH to investigate how the interplay of cosmological accretion, accretion onto the BH and the stabilizing effect of the BH can be important in determining the circumstances under which stars can form.

After the initial onset of star formation, we find that the radial extent of the star forming region remains relatively constant. Under the assumption of stars staying on circular orbits and not migrating in the disc, the process of forming stars increases the local surface density ($\Sigma_{\rm g} + \Sigma_{\star}$). This increases the self-gravity of the disc locally and decreases the effect of tidal forces on the gas. Removing the support from the tidal shear against gravitational collapse then leads to the further formation of stars in this same region. Following a short period beginning at the onset of star formation (while the stellar mass is still negligible), all subsequent star formation in the disc is largely confined to the region where stars have already formed. As stability increases in the presence of a massive BH, the radial extent of the region where stars can form narrows and the total SFR is reduced.

The radial extent of the region where stars can form in the model disc is small ($\sim 100$ pc) due to the disc properties at $z=6$, even in the absence of a BH. For the evolving model with a formation redshift at $z_{\rm i}=10$, we calculate the angular size of the stellar disc in the no BH case at $z=6$ to be $\theta< 0.02$ arcseconds and note this is less than the angular resolution of the James Webb Space Telescope (JWST), even at the shortest possible wavelengths. Resolved observations of such objects at this redshift would therefore by infeasible with current instruments.

The presence of a growing BH seed can greatly affect the star formation history of its host galaxy, even preventing the formation of stars entirely. Increasing the mass of the BH or the scale radius of the disc increases the stability of the disc, while increasing the disc mass decreases the stability. In the fiducial case, the disc becomes more unstable in the star forming region as the disc mass increases with the growth of the halo, resulting in SFR increasing with time. We find the sSFR in the model increases with higher BH mass and that the sSFR we calculate is higher than the observed median value at high redshift \citep{Stark2013}, particularly at times close to the onset of star formation. Our results suggest that systems hosting DCBHs should occupy the upper envelope of the sSFR distribution for any given stellar mass. Indeed, high sSFR galaxies could potential be used for the identification of DCBH hosts. As we evolve the model to lower redshifts, the discrepancy between the model sSFR and the observations decreases.

Increasing the accretion rate of the BH leads to an increase in the stability of the disc at a given time as the BH mass increases and the disc mass decreases. This can lead to a delay in the time where the disc first becomes unstable and forms stars. This delay in the onset of star formation is not only dependent on the BH growth rate and seed mass but also the growth of the disc and halo. As halo growth rates are higher at high redshift, the delay is also a function of the formation redshift of the BH. For a sufficiently high BH accretion rate and seed mass, the disc can be prevented from ever forming stars. At the lowest halo growth rates and high BH accretion rates, even models with early formation times have no stars forming. Such a low halo growth rate is typical of satellite galaxies \citep[see, e.g.][]{DeLucia2012}. This suggests the chance of a SMBH forming with no stellar disc counterpart is more likely in satellite galaxies. Indeed, this would also occur if an in-fall event were to occur prior to the onset of star formation.

We find that the halo in which a seed is born at $z=10$ is prevented from having significant star formation if the BH grows at the Eddington limit. If a seed BH is to grow at the rate required to increase in mass by $\gtrsim 3$ orders of magnitude between $z\sim10$ and $z\sim6$, star formation in its host is suppressed, placing such a system above the BH-stellar-mass relation. This suggests that DCBH galaxies will move towards the local BH-stellar mass relation via potential mergers with already evolved galaxies without massive BHs and not self-regulated co-evolution. Alternatively, this discrepancy can be resolved if either the formation of the DCBH is pushed to higher redshift ($z\sim20$) or if the evolution of the BH-galaxy system takes place in haloes with higher than average growth rates.

Though we do not model the feedback from the accreting BH we acknowledge that this would change the star formation and BH growth histories \citep{Schawinski2006, Latif2018}. BH feedback would heat and eject gas in the disc, acting to stabilise it, reducing the star formation rate in the model. The process of stabilising the disc through BH feedback would complement the gravitationally stabilising effect of the BH, delaying the onset of star formation further and decreasing the area of the disc able to form stars. This does not take into account the inclusion of ``positive feedback'' \citep{Gaibler2012}, where the inducing of star formation through jets leads to an increase in the SFR. However, this induced star formation would take place at large radii, meaning the inner region close to the BH would still be void of stars.

\section*{Acknowledgements}
\defcitealias{Astropy2013}{Astropy Collaboration, 2013}
DSE acknowledges the financial support of the Science
and Technology Facilities Council through a studentship
award. The Authors thank the reviewer for their helpful comments and suggestions. The Authors would also like to thank Tilman Hartwig, Bhaskar Agarwal and Marta Volonteri for their useful discussions and comments. This research made use of Astropy, a community-developed core Python package for Astronomy \citepalias{Astropy2013}.





\bibliographystyle{mnras}
\bibliography{librarybib} 



\appendix

\section{Change in Stability with Disc Scale Radius} 
\label{app:rd}

\begin{figure}
	\includegraphics[width=\columnwidth]{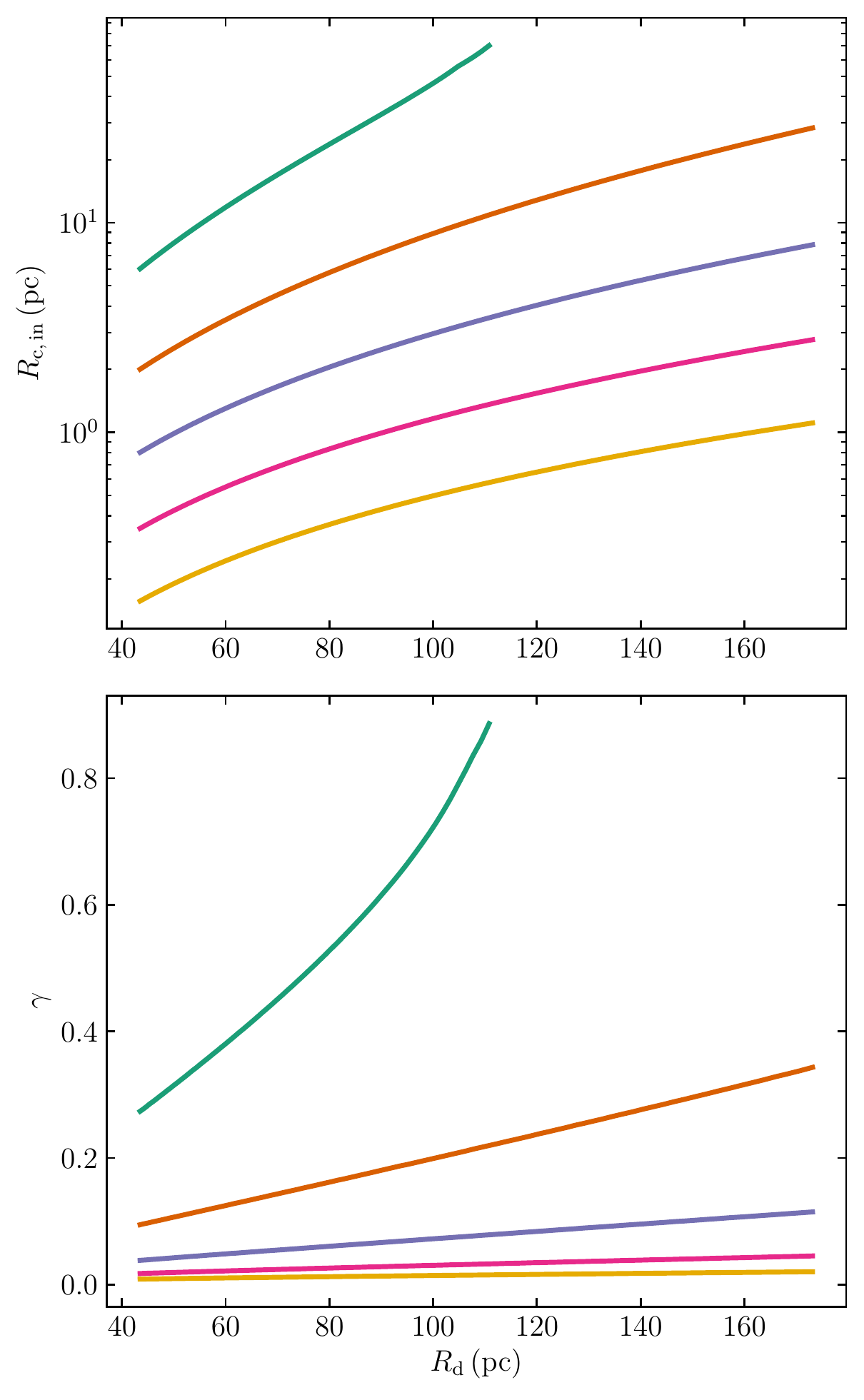}
	\caption{How the stability of the disc varies as a function of disc scale radius for different disc masses. The top panel shows the inner critical radius and the bottom shows that stable fraction of the disc. The lower limit to the range in $R_{\rm d}$ corresponds to $\lambda=0.025$ while the upper limit corresponds to $\lambda=0.1$. The disc masses are the same as in Figure~\ref{fig:RcvsMbh}: $10^8 \, M_{\odot}$ (green), $3.24 \times 10^8 \, M_{\odot}$ (orange), $10^9 \, M_{\odot}$ (purple), $3.24 \times 10^9 \, M_{\odot}$ (pink) and $10^{10} \, M_{\odot}$ (yellow).}
    \label{fig:vsrd}
\end{figure}

In the main body of this paper the disc scale radius is calculated using Equation~\ref{eq:Rd} with the mass of the system and the relevant redshift. Unless otherwise stated, the spin parameter of the halo is assumed as $\lambda = \bar{\lambda} = 0.05$ yet in nature $\lambda$ varies between haloes \citep{Mo1998}, resulting in a range of possible scale radii for the disc.

Figure~\ref{fig:vsrd} shows how varying $\lambda$, and therefore $R_{\rm d}$, changes the inner critical radius ($R_{\rm c, \,in}$) and stable fraction of the disc ($\gamma$). One can see from the figure that the stability of the disc is strongly sensitive to $R_{\rm d}$, particularly in the lowest mass case (the green line).  Increasing $R_{\rm d}$ decreases the surface density of the disc, raising the entire $Q_*$ profile, which leads to an increase in inner critical radius and stable fraction. In the lowest mass case, doubling $R_{\rm d}\sim43.3$ pc to $R_{\rm d}\sim86.5$ pc increases $R_{\rm c, \,in}$ by a factor of $\sim$5 and roughly doubles the value of $\gamma$.

As increasing $R_{\rm d}$ or $M_{\bullet}$ each result in an increase in the stable fraction of the disc the significance of the stabilising effect of the BH will vary with halo as the spin parameter varies and the resulting size of proto-galaxies vary.



\bsp	
\label{lastpage}
\end{document}